\documentclass[preprint,12pt]{elsarticle}

\usepackage{graphicx}
\usepackage{amssymb}
\usepackage{lineno}

\newtheorem{deff}{Definition}
\newtheorem{notei}{Notation}
\newtheorem{propo}{Proposition}

\newtheorem{lemm}{Lemma}

\newtheorem{rem}{Remark}

\usepackage{subfig}

\usepackage{xcolor}

\usepackage[ruled,vlined]{algorithm2e}

\usepackage{hyperref}
\hypersetup{
    colorlinks = false,
    linkbordercolor = {0.97254902 0.698039216 0.784313725}
}

\makeatletter
\newenvironment{algoproc}[1][]
  {\renewcommand{\algorithmcfname}{Procedure}%
   \begin{algorithm}[#1]
   \long\def\@caption##1[##2]##3{%
     \par
     \begingroup\@parboxrestore
     \if@minipage\@setminipage\fi
     \normalsize \@makecaption{\AlCapSty{\AlCapFnt\algorithmcfname}}{\ignorespaces ##3}%
     \par\endgroup
   }}
  {\end{algorithm}}
\makeatother

\SetCommentSty{mycommfont}

\usepackage{tikz}
\usepackage{PTSansNarrow}
\usepackage[T1]{fontenc}

\usetikzlibrary{matrix, positioning}

\usepackage[titletoc]{appendix}  

\usepackage{pgfplots}

\begin{document}

\begin{frontmatter}


\title{From Hashgraph to a Family of Atomic Broadcast Algorithms}



\author{Trafim Lasy\fnref{label1}}
\fntext[label1]{About author: Graduated from Ecole Normale Superieure (Paris). Received PhD in mathematics at Paris Diderot University in 2012.}
\ead{trafim.lasy@gmail.com}

\begin{abstract}
The goal of this article is to extend the ideas concerning asynchronous Byzantine Fault Tolerant (or BFT) consensus algorithm given in the work of Bracha, Toueg \cite{Bracha:1985:ACB:4221.214134} and Baird's Hashgraph consensus \cite{Baird:2016:1}. We propose a family of atomic broadcast algorithms, which Hashgraph consensus is closely related to. We also do preliminary comparative algorithm speed analysis which shows that some members of the family seriously outperform Hashgraph consensus. These algorithms can also be readily used as a base of proof-of-stake consensuses. In appendix we provide an extension of Hashgraph gossip protocol, which efficiently handles byzantine fault information exchange between nodes.
\end{abstract}


\end{frontmatter}

\newpage

\tableofcontents

\section{Introduction} \label{intro}
The problem of Atomic Broadcast is well-known in the theory of distributed systems. Informally, \textit{Atomic Broadcast} is a protocol allowing a set of processes (or nodes) to broadcast messages in such a way that processes agree on the set of messages they deliver and the order of messages delivered. Such protocols can be used in cryptocurrency and blockchain domains, which recently gained much attention. One such example is Hashgraph consensus algorithm. In the original paper \cite{Baird:2016:1} the consensus algorithm stands a bit isolated from a 40+ year theory of distributed algorithms. One of my original goals was to answer many ``why it is done this way and not the other one?'' questions, which arose while reading the paper. My research not just allowed me to answer most of them but also revealed one BFT consensus protocol Hashgraph consensus relates to. Namely, I showed that Hashgraph \textit{decideFame} procedure is closely connected to consensus protocol for the malicious case proposed by Bracha and Toueg \cite{Bracha:1985:ACB:4221.214134} in 1985. It is worth mentioning that their paper also contains the famous result that more than 2/3 of processes should be correct in order to reach the consensus with probability 1. My study also allowed to construct a family of atomic broadcast algorithms generalizing Hashgraph consensus ideas. Some of these algorithms are possibly faster than and as safe as Hashgraph consensus. By ``possibly faster'' I mean that I did a comparative speed analysis on a big set of message exchange and process fault scenarios showing their superiority. Mathematical proof is an open question and the answer will probably depend on the message exchange probability model chosen.  

The remainder of this article is organized as follows:
\begin{itemize}
 \item In Section \ref{net_model_main_def} we give the network model and all necessary definitions from distributed algorithm theory and Hashgraph paper. 
 \item Section \ref{hash_classic} is devoted to the description of Hashgraph consensus, in particular its core \textit{decideFame} procedure and its connection to classic theory of distributed algorithms. 
 \item In Section \ref{section.BVC} we provide a description of atomic broadcast family of algorithms, which we call BVC family. We give some examples and show the appropriateness of algorithms. 
 \item We define speed using latency metrics and do the corresponding analysis of Hashgraph consensus and several algorithms from BVC family in Section \ref{speed_comparison}.
 \item In Section \ref{section.practical_future} we briefly remind practical use cases of atomic broadcast algorithms discussed in the paper. We also put final remarks, future work thoughts and acknowledgement there. 
 \item In the Appendix \ref{our_gossip_proto} we propose an optimization for Hashgraph gossip protocol adding the rules for efficient byzantine (malicious) fault information exchange between nodes.
\end{itemize}

\section{Network Model and Main Definitions} \label{net_model_main_def}
We start this section with the network model part then proceed with the main definitions.
\subsection{Network Model Description and Justification} \label{net_model}
Throughout this paper we treat the words ``process'' and ``node'' as synonyms. We stick to a variant of standard \textit{asynchronous network model} consisting of $n$ interconnected nodes whose messages can be delayed arbitrarily long and some of them may even be lost. Part of the nodes can be faulty (they are also called ``malicious'' of ``byzantine'') and can make a ``coordinated attack'' (loosing/delaying messages, sending malformed ones etc.) on the rest of the nodes (which act according to protocol rules and are called ``honest'' or ``correct'') in order to interrupt the correct work of the whole system. By ``correct work'' be mean providing a fault-tolerant distributed database service via consistent ordering of messages bearing database updates/transactions. This is equivalent to atomic broadcast mentioned in the introduction. We also require that more than $2n/3$ nodes are honest and for any number $r$ there will an honest node with a-round greater than $r$ (below we define what a-round is and provide the justification for these requirements). 

Atomic broadcast is equivalent to consensus protocol with Byzantine faults (cf. Proposition 7 in \cite{Milosevic:2011:RAB:2085039.2085372}, for example) which can be defined as follows:
\begin{deff}[Consensus]
 In the \textbf{consensus protocol} every node has an initial value from a set $V$ and has eventually to irrevocably decide on a value from V. The following properties should be satisfied:
 \begin{itemize}
  \item Agreement. All honest nodes decide on the same value.
  \item Validity. If all honest nodes start with the same value $v$, then all honest nodes decide on $v$.
  \item Termination. All honest nodes eventually decide.
 \end{itemize}
\end{deff}
Famous FLP theorem \cite{Fischer:1985:IDC:3149.214121} implies that consensus protocol is impossible even with one faulty node. This motivated Bracha and Toueg \cite{Bracha:1985:ACB:4221.214134} to work with consensus where the \textit{expected termination} time is finite. In \cite{Bracha:1987:ABA:36888.36891} Bracha also defined this property using the notion of \textit{asynchronous rounds}: 
\begin{deff}[Bracha] \label{bracha}
 In a distributed system with maximum $f$ faulty processes:
 \begin{itemize}
  \item We define inductively \textbf{asynchronous rounds} (or simply \textbf{a-rounds}) as follows: in each a-round, every node sends messages to all others, waits for only $n - f$ messages of that a-round, and proceeds to the next a-round.
  \item  We say that consensus protocol has a \textbf{probabilistic termination} property if the probability that an honest node is undecided after $r$ a-rounds approaches zero as $r$ approaches infinity.
 \end{itemize}
\end{deff}
\begin{rem}
 Bracha and Toueg originally used the word ``round'', which we intentionally replaced by ``a-round'' in order to not confuse them with Hashgraph's rounds introduced later.
\end{rem}
Consensus protocol is possible with probabilistic termination. An already mentioned Bracha and Toueg result in \cite{Bracha:1985:ACB:4221.214134} states that the necessary condition for this is that the maximal possible number $f$ of faulty processes is less than $n/3$. In the same paper they have also proven that this was a sufficient condition assuming the so-called \textit{fair scheduler} property of the network and giving an example of consensus protocol, which we will describe in the next section. Bracha continued their work in \cite{Bracha:1987:ABA:36888.36891} removing that additional assumption using a \textit{coin toss} trick. As we see, the above results naturally justify all the properties of our network model.

\subsection{Main Definitions} \label{main_defs}
\begin{figure}[h!]
  \centering\includegraphics[width=0.3\linewidth]{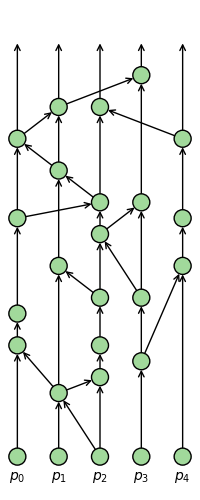}
  \caption{Hashgraph for 5 nodes. Vertices are events, edges - ``parent'' relation.}
  \label{fig:hashgraph-lamport}
\end{figure}

In this subsection we briefly remind main definitions from Baird's Hashgraph consensus article \cite{Baird:2016:1} adding comments and links to classic notions.
\begin{itemize}
 \item \textit{Gossip protocol} means that any node $p$ chooses another node $q$ at random and then $p$ sends (or ``gossips'') $q$ all of the information it knows so far. After that (or at the same time) $p$ repeats this procedure with a different node and so on. Node $q$ and all the other nodes repeatedly do the same.  
 \item \textit{Hashgraph} is the directed graph that represents the gossip history: who gossiped to whom, and in what order. Figure \ref{fig:hashgraph-lamport} contains an example of Hashgraph.   
 \item Vertices in Hashgraph  are called \textit{events} and stored in memory as a sequence of bytes, signed by its creator. Nodes should obey the following rule: If node $p$ gossiped to node $q$ then the latter must create an event $e$, which is signed by $q$ and contains the hashes of two other events: $q$'s last event and $p$'s last event prior to that gossip. 
 \item Each of the last two events mentioned in the previous definition is called \textit{parent} of the newly created event $e$. To distinguish them we will also call $q$'s last event \textit{self-parent} of $e$. As one can see the events of nodes that obey the above rule form a sequence.
 \item On creation node also puts it's local time into event as event's \textit{timestamp}.
\end{itemize}
\begin{rem}
It is worth pointing out that the notions presented above are not new. Term ``event'' is present in numerous works starting, probably, with the influential paper of Lamport \cite{Lamport:1978:TCO:359545.359563}. Hashgraph, in its turn, is an analogue of ``space-time diagram'' (cf. Lamport \cite{Lamport:1978:TCO:359545.359563}) or ``communication pattern'' (cf. Lynch \cite{Lynch:1996:DA:2821576}), both of which represent node message exchange history. 
\end{rem}

Noting that ``parent'' relation is represented by arrows (directed edges) in hashgraph we continue with definitions:
\begin{itemize}
 \item Informally, event $e$ \textit{follows} event $e'$ if $e'$ is a parent of an event, which is a parent of an event, which is .. and so on until $e$. We allow an event to follow itself and say that event $e'$ is an \textit{ancestor} of an event $e$ if $e$ follows $e'$. On hashgraph this means that there is a downward path from $e$ to $e'$. Mathematically, ``follows'' relation is the transitive closure of the inverse to the ``parent'' one (similar definitions can be found, for example, in \cite{Lamport:1978:TCO:359545.359563} and \cite{Moser:1999:BTO:300547.300555}).     
 \item As it was mentioned in the above definition of \textit{event}, honest node should create events one by one and always put the hash of its previous event in the next one. This implies that if we take any two events created by the same honest node then one of them is an ancestor of the other. Faulty node can violate this rule and on creation put a hash of an event different from the previous one. This produces pairs of events created by the same node neither of which is an ancestor of the other one. We call any such pair $(e, e')$ a \textit{fork} and say that $e'$ is a \textit{fork} of event $e$.  
 \item Key notions in Hashgraph consensus are ``seeing'' and ``strongly seeing''. Event $e$ \textit{sees} event $e'$ is $e$ follows $e'$ and follows no fork by the creator of $e'$. An event $e$ \textit{strongly sees} an event $e'$ if $e$ sees $e'$ and $e$ sees more than $2n/3$ events created by different nodes, each of which sees $e'$.  
 \item Such core notion as \textit{Virtual voting} can be informally described the following way: every node has its version of the hashgraph, so node $p$ can calculate, using hashgraph's graph structure, what vote node $q$ would have sent to it if they had been running a traditional consensus protocol with Byzantine faults that involved sending votes. Since the hashgraph alone is sufficient no votes need to be sent, and one can say that nodes \textit{vote virtually}. We'll elaborate on Virtual voting in the next section.
 \item Using the above definitions Baird inductively gives the notion of \textit{round $R$ witness} (or $R$-witness for short): All starting events for each node are round $1$ witnesses. An event is $R+1$-witness for the given node if it is node's earliest event (in the sense of ``follows'' relation) which strongly sees more than $2n/3$ of $R$-witnesses.
 \item Event \textit{rounds} are naturally derived from witnesses: an event has \textit{round $R$} if it follows an $R$-witness and doesn't follow any $R+1$-witness. 
\end{itemize}

\section{Hashgraph Consensus and Classic Theory} \label{hash_classic}
This section is devoted to the connection of Hashgraph consensus and classic distributed algorithm theory. Hashgraph algorithm \cite{Baird:2016:1} can be presented via mathematical pseudocode as follows:

\begin{algoproc}[H]
  \DontPrintSemicolon
  run two loops in parallel:\;
  \While{TRUE} {
    gossip all known events to a random node
  }
   \While{TRUE} {
    receive a gossip\;
    create a new event\;
    \textbf{call} divideRounds\;
    \textbf{call} decideFame\;
    \textbf{call} findOrder
  }
  \caption{Hashgraph consensus algorithm}
\end{algoproc}

As one can see, algorithm de-facto consists of three procedures. First, \textit{divideRounds} procedure is just event round and witness calculation and can be strictly described as:

\begin{algoproc}[H]
  \DontPrintSemicolon
  \For{each new event $x$} {
    $r \leftarrow$ max round of parents of $x$ (or $1$ if none exist)\;      
    \uIf{$x$ can strongly see more than $2n/3$ round $r$ witnesses} {
	$x.round \leftarrow r + 1$
    }
    \Else{
	$x.round \leftarrow r$
    }
    $x.witness \leftarrow$ ($x$ has no self-parent) or ($x.round > $ round of $x$'s self-parent)
  }
  \caption{divideRounds}
\end{algoproc}

Third, \textit{findOrder} procedure is also quite simple and bears the rules for adding events to event total order. We'll briefly discuss it at the end of this section. Most interesting is the second, \textit{decideFame} procedure. We'll show that under gossip protocol it's main part is similar to the f-resilient consensus protocol for the malicious case presented in Bracha and Toueg \cite{Bracha:1985:ACB:4221.214134}. For this we will need an important definition we intentionally kept until now:
\begin{deff}[Baird. Famous witness. Informally] 
 A witness is defined to be \textbf{famous} if the hashgraph structure shows that most nodes received it fairly soon after it was created.  
\end{deff}
Words ``fairly soon'' will become clear after we present \textit{decideFame} as a mathematical pseudocode. Note that we present the core of this procedure, removing rather artificial ``coin flip'' part of it (cf. \cite{Baird:2016:1} for the whole version). The sole goal of coin tosses, flips, etc., is to guarantee that consensus will be reached with probability one even in the ``worst''  and unlikely scenarios. As it was mentioned during the network description, such tricks are rather standard (cf. for example Protocol 2 in \cite{Bracha:1987:ABA:36888.36891}) in distributed algorithm theory. 

\begin{algoproc}[H]
  \DontPrintSemicolon
  $Y \leftarrow$ set of newly added/created witnesses on the gossip receipt, ordered by ``follows'' relation, earliest going first\;
  \For{$y$ in $Y$} {
    $y.famous \leftarrow Undecided$\tcp*[r]{initializing witness fame}      
    $X \leftarrow$ set of all witnesses with undecided fame and of round lesser than $y.round$\;
    \For{$x$ in $X$} {
      $d \leftarrow y.round - x.round$\tcp*[r]{$d > 0$}
      \uIf{$d = 1$} {
	$y.vote_x \leftarrow$ can $y$ see $x$ ?\tcp*[r]{$vote_x$: vote $x$ is famous or not}
      }
      \Else(\tcp*[f]{$d\geq 2$}){
	  $S \leftarrow$ set of all round $(y.round - 1)$ witnesses that $y$ can strongly see\;
	  $v \leftarrow$ majority vote in $S$ for whether $x$ is famous or not (is TRUE for a tie)\;
	  $t \leftarrow$ number of events in $S$ with a $vote_x$ equal to $v$\;
	  $y.vote_x \leftarrow v$ \tcp*[r]{voting}	  
	  \If{$t > 2n/3$} {
	    $x.famous \leftarrow v$\tcp*[r]{deciding}
	  }
      }
    }
  }
  \caption{decideFame (no coin flip)}
\end{algoproc}

As Baird fairly notes, this procedure consists of parallel BFT binary consensuses on whether a witness is famous or not. This is done via virtual voting in the sense that no special voting messages are being sent and just the hashgraph structure is being used. As promised, we explain in detail this statement and show that BFT binary consensuses above are similar to the one in Bracha and Toueg \cite{Bracha:1985:ACB:4221.214134}.

Fame decision for the given $R$-witness is converted to the binary consensus as follows: each node that created an $(R+1)$-witness participates in consensus with initial value $0$ if the latter witness sees that round $R$ witness, or $1$ if it doesn't. Note that \textit{divideRounds} procedure does not guarantee that every active node will have round $R+1$ witness (by ``active'' we mean that node creates arbitrarily large round witnesses as time goes on). Indeed, sometimes an active node can pass directly from $R$-witness to $(R+2)$-witness. At the same time  classic consensus problem requires every node to start with some value. In order to overcome this technical issue we'll show the similarity of \textit{decideFame} consensus and Bracha-Toueg consensus under ``round completion'' procedure: 
\begin{deff} \label{round compl}
 By \textbf{round completion} we mean the following addition to round definition: when a node creates $(R+K)$-witness ($K > 1$) right after $R$-witness then we consider that $(R+K)$-witness to be also round $(R+1), \ldots, (R+K-1)$ witness. 
\end{deff}
This addition clearly guarantees that every active node will have an $(R+1)$-witness.     

Bracha and Toueg $f$-resilient consensus protocol for the malicious case in \cite{Bracha:1985:ACB:4221.214134} is given as \textbf{algorithm \ref{algo:1}} pseudocode. 

\begin{algorithm}[hp]
  \DontPrintSemicolon
  \SetAlgoRefName{1}
  \While{TRUE} {
    $message\_count \leftarrow 0$; $phaseno \leftarrow 0$; $echo\_count \leftarrow 0$\;
    \ForAll{$1\leq q \leq n$} {
      send(q, (initial, p, value, phaseno))
    }
    \While{message\_count(0) + message\_count(1) $< n - f$} {
      receive(msg)\;
      \If{it is the first message received from the sender with these values of msg.type, msg.from and msg.phaseno} {
	\uIf{msg.type = initial} {
	  \ForAll{$1\leq q \leq n$} { 
	    send(q, (echo, msg.from, msg.value, msg.phaseno))
	  }
	}
	\uElseIf{msg.type = echo and msg.phaseno = phaseno} {
	  echo\_count(msg.from, msg.value) += 1\;
	  \If{$echo\_count(msg.from, msg.value) = \lfloor(n + f)/2\rfloor + 1$} {
	    message\_count(msg.value) += 1
	  }
	}
	\ElseIf{msg.type = echo and msg.phaseno $>$ phaseno} {
	  send(p, msg)
	}
      }
    }
    \uIf{message\_count(1) $>$ message\_count(0)} {
      value $\leftarrow 1$
    } \Else {
      value $\leftarrow 0$
    }
    \If{there is $i$ such that $message\_count(i) > (n + f)/2$} {
      $d_p \leftarrow i$ \tcp*[r]{deciding}
    }
    phaseno += 1\;
  }
  \caption{Node $p$ : f-consensus. Value is either $0$ or $1$ (initialized with $p$'s starting value). Phaseno: integer (analogue of rounds). Message\textunderscore count: tuple, representing given value message counts. Echo\textunderscore count: $n\times 2$ array representing echoes of messages broadcasted by nodes. Message consists of type: ('initial', 'echo'), from (node id), value, phaseno.
  }
  \label{algo:1}
\end{algorithm}

Informally, this consensus can be described as follows: In order to decide which value to choose, processes repeatedly broadcast their current phase number and a binary value. They use phases to overcome misleading messages from the malicious processes, and accept a value from a process if enough other processes confirmed (echoed) that a given value was indeed broadcasted by this process. After receiving enough values, process chooses the new one by majority principle and passes to the next phase. A process decides $i$ if it accepts more than $(n + f)/2$ messages with value $i$ (here $f$ is a maximal possible number of malicious processes). 

\begin{figure}[h!]
 \begin{minipage}{.5\linewidth}
  \centering
  \subfloat[]{\label{fig:vv0}\includegraphics[scale=.8]{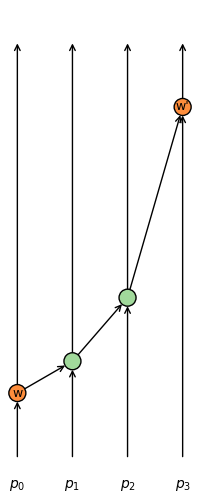}}
 \end{minipage}
 \begin{minipage}{.5\linewidth}
  \centering
  \subfloat[]{\label{fig:vv1}\includegraphics[scale=.8]{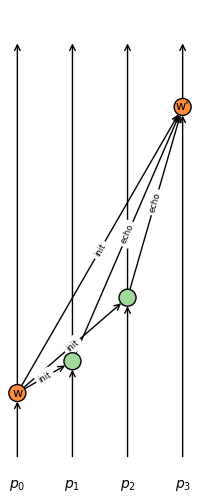}}
 \end{minipage}\par\medskip
 \caption{Virtual voting example. On both figures $w$ is a witness and $w'$ is next round witness. The fact that $w'$ sees nodes $p_1$ and $p_2$ events that see $w$ on figure (\ref{fig:vv0}) can be interpreted as a set of $init$, $echo$ messages on figure (\ref{fig:vv1}). 
 }
 \label{fig:virtual_voting}
\end{figure}

In order to prove the main result of this section we'll need the following definition:
\begin{deff} \label{relaxed_ss}
 We define \textbf{relaxed strongly seeing} the same way ``strongly seeing'' is defined except that we replace ``more than $2n/3$'' by ``more than $(n + f)/2$'' ($f = \lfloor(n - 1)/3\rfloor$)  events created by different nodes. 
\end{deff}
\begin{rem}
 ``Relaxed strongly seeing'' and ``strongly seeing'' definitions are the same when $n$ is not divisible by $3$. ``Strongly seeing'' is stricter when $3$ divides $n$.
\end{rem}
Now we are ready to prove the following proposition:
\begin{propo} \label{hg_bt_sim}
 With relaxed strongly seeing and round completion procedure BFT binary consensuses in {\normalfont decideFame} procedure follow from enriched Bracha-Toueg $f$-resilient consensus protocol for the malicious case with $f = \lfloor(n - 1)/3\rfloor$.
\end{propo} 
\textit{Proof.} The proof is quite straightforward and starts with the following interpretations:
\begin{itemize}
 \item {[Fame as consensus]} As already mentioned, fame decision for the given $R$-witness $w$ can be considered as consensus problem where each active node starts with initial value $0$ if its $(R+1)$-witness sees $w$ and $1$ if it doesn't.
 \item {[Virtual voting]} The fact that node $p$'s $(R+K+1)$-witness sees node $q$'s event that, in its turn, sees node $r$'s $(R+K)$-witness $w_r$ ($K \geq 1$) can be interpreted as situation when $r$ has sent ($initial$, $r$, $w_r.vote$, $K-1$) message to $q$ and the latter echoed this message as ($echo$, $r$, $w_r.vote$, $K-1$) to node $p$. See an example on figure \ref{fig:virtual_voting}.
 \item {[Gossip protocol]} Taking into account the above interpretation, we treat gossip receipt by a process in Hashgraph consensus as a batch of $receive(msg)$ virtual voting interpretations in Bracha-Toueg protocol, and this batch should be processed without exiting the \textit{inner while loop} in algorithm \ref{algo:1}.    
 \item {[Phase-to-round correspondence]} Finally, despite having very similar inductive definitions phases and rounds are not in general the same: phases are based on binary value count, whereas rounds on witness hash counts. In order to completely equalize them message content in Bracha-Toueg protocol should be \textit{enriched} as follows: whenever a process passes to the next phase it generates a \textit{number}, which is broadcasted and echoed alongside with \textit{value}. If $echo\_count$ takes into account tuples (\textit{number, value}) instead of just \textit{value} then phases become equal to rounds when \textit{number} generated by a process is just the latest witness hash. Indeed, one can easily verify that in this case: 
 \begin{itemize}
  \item the fact that the latest node $p$ event starts to strongly see (in relaxed way) node $q$ witness of the previous round is equivalent to  
  $$echo\_count(q, (number, value)) \geq \lfloor(n + f)/2\rfloor + 1$$ 
  for $f = \lfloor(n - 1)/3\rfloor$.  
  \item the fact that newly created node $p$ event is the next round witness is equivalent to 
  $$message\_count(0) + message\_count(1) \geq n - f.$$  
 \end{itemize}
\end{itemize}

In order to finish the proof we need to show that decision procedures in \textit{decideFame} and Bracha-Toueg $f$-resilient consensuses lead to the same results. Recall that \textit{decideFame} consensuses require that more than $2n/3$ witnesses strongly seen by next round witness have the same vote $v$, while Bracha-Toueg consensus requires (under interpretation) that more than $(n + f)/2$ witnesses, which are strongly seen by newly created next round process $p$ witness, have $v$ as a vote. It is not hard to see that both these conditions imply that all next round witnesses will vote $v$. Thus the decisions are equivalent and the proof is finished. $\square$ 

We presenting the last, \textit{findOrder} procedure not only for completeness reasons. In the next chapter we will present several speed enhancements for this procedure. We start with the following definitions from \cite{Baird:2016:1}: 
\begin{deff}
 \begin{itemize}
  \item Once all the witnesses in round $r$ have their fame decided, find the set of famous witnesses in that round, then remove from that set any famous witness that has the same creator as any other in that set. The remaining famous witnesses are called \textbf{unique famous witnesses}. 
  \item The \textbf{round received number} (or \textbf{round received}) of an event $x$ is defined to be the first round where all unique famous witnesses follow $x$.
  \item We call an event \textbf{self-ancestor} of event $e$ if $e$ follows this event and they have the same creator. 
 \end{itemize}
\end{deff}
The \textit{findOrder} procedure is presented in \cite{Baird:2016:1} as the following preudocode:
\begin{algoproc}[H]
  \DontPrintSemicolon
  \ForAll{events $x$ with undefined $x.roundReceived$} {
    \If{there is a round $r$ such that there is no witness $y$ in or before round r that has $y.famous = Undecided$ \\
    {\normalfont \textbf{and}} $x$ is an ancestor of every round $r$ unique famous witness \\
    {\normalfont \textbf{and}} this is not true of any round earlier than $r$} {
      $x.roundReceived \leftarrow r$ \;
      $s \leftarrow$ set of each event $z$ such that $z$ is a self-ancestor of a round $r$ unique famous witness, and $x$ is an ancestor of $z$ but not of the self-parent of $z$ \;
      $x.consensusTimestamp \leftarrow$ median of the timestamps of all the events in $s$
    }
  }
  \Return{all events with defined $roundReceived$, sorted by $roundReceived$, then ties sorted by $consensusTimestamp$, then by whitened signature}
  \caption{findOrder}
\end{algoproc}
As one can see, once a next round unique famous witnesses are defined there's a possibility that one or several events got their $consensusTimestamp$, are considered to be committed and added to event total order. Baird proves that this order is invariant among nodes, thus proving that Hashgraph consensus is atomic broadcast algorithm. Having described Hashgraph consensus we can pass to the next section concerning its possible generalizations and enhancements.  

\section{BVC Atomic Broadcast Family}  \label{section.BVC}
Recall from the introduction that atomic broadcast algorithms provide a way for nodes to consistently order messages or events they create and send to each other. As usual in literature, order consistency means that if any honest node determines that $e$ is the $i$'th event of the total order, then no honest node determines that $e'$ is the $i$'th event, where $e'\neq e$ (cf. \cite{Moser:1999:BTO:300547.300555} for example). Leemon Baird notes in \cite{Baird:2016:1} that to eventually order all events it is enough to accomplish two tasks: 
\begin{itemize}
 \item consistently define the event subset consisting of famous witnesses, using fame voting
 \item derive the total order for all events afterwards, using procedure \textit{findOrder}    
\end{itemize}
In this section we present a family of atomic broadcast algorithms, elaborating on generalizations and improvements for both of these tasks. We will call this family \textbf{BVC Family} since it will be built on three notions: Base, Voting and Consensus layers. 

\subsection{Event Subset Choice Task}
One of the first questions, which arise looking at how Hashgraph consensus solves the first task, is ``Why should we start with witnesses as candidates for being famous?''. Witnesses of different rounds form a sequence of layers starting with the layer of first round witnesses, then second round witnesses and so on.  This simple observation induces a question: ``Can we choose another sequence of layers as fame candidates?''. The answer is ``yes''. We will show that there are many possibilities for such layer sequences. 

Another question is ``Why particular witness fame is decided by the layer of next (or next-but-one) round witnesses?''. Here, again the answer is ``there are many possibilities to choose this layer''. Natural condition for this layer is that it should contain sufficiently events in order to conduct fame voting procedure afterwards (be it BFT consensuses from the previous section or any other binary consensus algorithm). 
\begin{deff}
 Elements of fame candidate layer sequence are called \textbf{base layers}. Corresponding layers of voting events are called \textbf{voting layers}. 
\end{deff}
Appropriate definitions of base and voting layers, combined with fame voting BFT consensus, allow to construct new family of algorithms. Here word ``appropriate'' bears the following meaning:
\begin{deff}[Appropriateness] \label{approp_def}
 We say that base layers, voting layers and fame consensus definitions are \textbf{appropriate} if the following conditions are satisfied:
 \begin{enumerate}
  \item The fact that an event belongs to a particular base or voting layer is calculated using only event itself and its ancestors. This guarantees that all honest nodes will have consistent base and voting layers.
  \item If an event is known by more than $2f$ nodes then eventually it will be an ancestor of some famous event from some base layer (now and further $f = \lfloor(n - 1)/3\rfloor$ is the  upper theoretical bound for the faulty nodes number).
 \end{enumerate} 
\end{deff}
\begin{rem}[Atomic broadcast]
 As we will see in ``total order'' part of this section, second appropriateness condition implies that every event known by more than $2f$ nodes will be committed (added to the total order) with probability $1$. And this order will be BFT and consistent across the honest nodes implying the {\normalfont atomic broadcast} property.
\end{rem}

Now we provide some examples of base and voting layers construction. We also give a version of fame voting consensus. This will show how rich our atomic broadcast family is.

\subsection{Base Layer Examples} \label{base_layers}

\begin{figure}[h!]
 \begin{minipage}{.32\linewidth}
  \centering
  \subfloat[]{\label{fig:base_ex_a}\includegraphics[scale=.7]{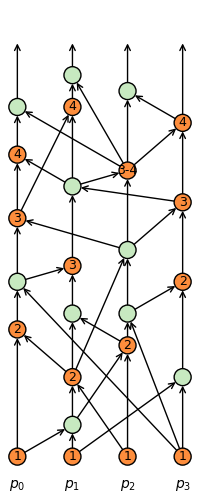}}
 \end{minipage}
 \begin{minipage}{.32\linewidth}
  \centering
  \subfloat[]{\label{fig:base_ex_b}\includegraphics[scale=.7]{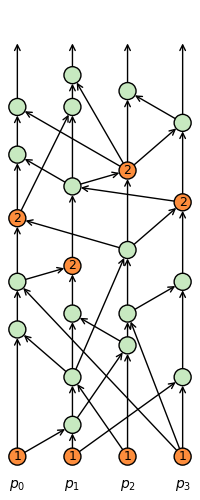}} 
 \end{minipage}
 \begin{minipage}{.32\linewidth}
  \centering
  \subfloat[]{\label{fig:base_ex_c}\includegraphics[scale=.7]{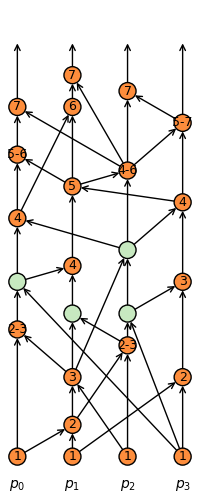}}
 \end{minipage}\par\medskip
 \caption{On all figures we mark events belonging to some base layer with orange color and by light green the rest events. Event belonging to $k$\textsuperscript{th} base layer is marked by label `k'. We also mark with label `x-y' events simultaneously belonging to $x$\textsuperscript{th}, $(x+1)$\textsuperscript{th}, \ldots, up to $y$\textsuperscript{th} base layers. Figure (\ref{fig:base_ex_a}) represents an example of a-base layer, figure (\ref{fig:base_ex_b}) is an example of s-base layers (or Hashgraph witnesses), and figure (\ref{fig:base_ex_c}) gives an example of $C'(1, 4)$-base layers. All examples have the same gossip history. Since $n=4$, the maximal number $f$ of malicious processes is $1$ and $n-f = 3$.}
 \label{fig:base_ex}
\end{figure}

Recall that in network model subsection \ref{net_model} we had Bracha's asynchronous rounds definition (cf. a-round definition \ref{bracha}). It gives rise to the following ``asynchronous'' base layer construction (which is called a-base layer, for short):
\begin{deff}[A-base layer] \label{def.a-base} 
We define inductively \textbf{a-base layers} as follows:
 \begin{itemize}
  \item Different node starting events form first a-base layer.
  \item For each node: earliest node's events that follow at least $n-f$ created by different nodes events from $(k-1)$\textsuperscript{th} a-base layer belong to $k$\textsuperscript{th} a-base layer (word ``earliest'' means that there's no other events with that property followed by the given event, except itself).  
 \end{itemize}
\end{deff}
\begin{rem}
 A-base layer definition implies that an event can be a member of several consecutive a-base layers similarly to round completion procedure (cf. definition \ref{round compl} in section \ref{hash_classic}). An example is given on figure (\ref{fig:base_ex_a}).
\end{rem}
For completeness and uniformity, we give the following example of base layer, which is, obviously, used by Hashgraph consensus algorithm:
\begin{deff}[S-base layer] \label{def.s-base} 
 We define $k$\textsuperscript{th} \textbf{s-base layer} (``s'' due to ``strongly see'' phrase) as the set consisting of all $k$ round witnesses in the sense of Hashgraph (cf. figure (\ref{fig:base_ex_b})).
\end{deff}
To give next example we need the following important definitions:
\begin{deff}[Clear following]
 We say that an event $e$ \textbf{clearly follows} and event $e'$ if $e$ follows $e'$ and follows no fork of $e'$. 
\end{deff}
\begin{rem}
 ``Clear following'' is weaker than ``seeing'': e.g. an event $e$ follows a fork $(e_1, e_2)$ that follows event $e'$. In this case $e$ still clearly follows $e'$ but doesn't see it anymore!
\end{rem} 
\begin{deff}[Strongly following]
 We say that event $e$ \textbf{strongly follows} event $e'$ if $e$ clearly follows $e'$ and $e$ follows (and not necessarily sees !) more than $(n + f)/2$ events created by different nodes, each of which clearly follows $e'$ (again, not necessarily sees). 
\end{deff}
Replacing ``following'' by ``strongly following'' in a-base layer definition, we get: 
\begin{deff}[S$'$-base layer] \label{def.s'-base} 
 We define inductively \textbf{$s'$-base layers} as follows:
 \begin{itemize}
  \item Different node starting events form first $s'$-base layer.
  \item For each node: earliest node's events that strongly follow at least $n-f$ created by different nodes events from $(k-1)$\textsuperscript{th} $s'$-base layer belong to $k$\textsuperscript{th} $s'$-base layer.  
 \end{itemize}
\end{deff}
\begin{rem}
 Honest nodes can have only one event in the given a-base, s-base or $s'$-base layers, while malicious nodes, when start forking, can have several.  
\end{rem}
Looking at a-base layer definition one can ask ``Why can't be replace `following $n-f$ elements' by `following smaller number of elements', say $2$?''. The answer is ``Yes, we can, but with a small adjustment''. If we require that next base layer elements follow a small number of previous layer ones, then some group of malicious nodes can quickly gossip between each other, making far higher base layers then the rest of the nodes. This group then gossips its events to other nodes, making sure that these gossiped events will be famous. In order to counter such de-facto malicious behavior one can demand that at least from time to time next layer elements should be assigned using $n-f$ different node events. These reasonings give rise to the following   
\begin{deff}[C(a, b)-base layer] \label{def.c-base} 
 Let $1 < a \leq n-f$ and $b > 0$ be some integers. \textbf{$\mathbf{C(a, b)}$-base layers} are defined as follows:
 \begin{itemize}
  \item All node starting events are put into $C(a, b)$-base layer $1$.
  \item Event belongs to $k$\textsuperscript{th} $C(a, b)$-base layer if:
  \begin{itemize}
   \item it is node's earliest event that follows at least \textbf{a} created by different nodes events from $(k-1)$\textsuperscript{th} $C(a, b)$-base layer for $k$ \underline{not} divisible by $b$
   \item it is node's earliest event that follows at least $\mathbf{n-f}$ created by different nodes events from $(k-1)$\textsuperscript{th} $C(a, b)$-base layer for $k$ divisible by $b$
  \end{itemize}
 \end{itemize}
\end{deff}
 We see that in the above definition each $b$\textsuperscript{th} layer requires $n-f$ different nodes events. This property will also be used later when we will prove ``appropriateness'' properties.
\begin{deff}[C$'$(a, b)-base layer] \label{def.c'-base} 
 If in $C(a, b)$-base layer definition we require instead that $k$\textsuperscript{th} base layer event $e$ is node's earliest event that follows at least \textbf{a} \underline{different from $e$} and  created by different nodes events from $(k-1)$\textsuperscript{th} base layer for $k$ not divisible by $b$, then we get the definition of \textbf{$\mathbf{C'(a, b)}$-base layer}. Figure (\ref{fig:base_ex_c}) is a particular example. 
\end{deff}
\begin{rem}
 One can see that $C'(a, b)$-base layer definition is also valid for $a=1$. In this case $C'(1, b)$-base layers are tightly connected with the completion of the classic {\normalfont Lamport time} notion. Cases $C(2, b)$ and $C'(2, b)$ also provide interesting examples of base layers.
\end{rem}
\begin{rem}
 It is clear that all the above base layer definitions are based on events and their ancestors only, thus satisfying the first property of appropriateness definition \ref{approp_def}. 
\end{rem}
We finish base layer examples by generalizing them and giving a generic one:
\begin{deff}[G-base layer] \label{g_base_def}
 We define a \textbf{generic base layer} (or \textbf{g-base layer}) as follows: 
 \begin{enumerate}
  \item Any node starting events are assigned to g-base layer $1$.
  \item We form higher g-base layers using an inductive rule: $k$\textsuperscript{th} g-base layer is built on $(k-1)$\textsuperscript{th}.
  \item The above rule satisfies the first appropriateness definition \ref{approp_def} condition. 
  \item The above rule ensures that for any event, followed by more than  $2f$ different node events, there will be number $k_0$, such that this event will eventually be followed by any $k$\textsuperscript{th} g-base layer with $k > k_0$. 
  \item Any g-base layer contains at least $n-f$ eventually strongly seen elements.  
 \end{enumerate}
\end{deff}
\begin{lemm}
 A-base, s-base, $s'$-base, $C(a,b)$ and $C'(a,b)$-base layers are particular cases of g-base layers.    
\end{lemm}
\textit{Proof.} Three first g-base layer properties are obvious. The fifth one follows from our network model property that a-rounds are not bounded (cf. \ref{net_model}). We prove the hardest, the fourth g-base property. All mentioned in lemma statement base layers have a common feature: at least subsequence of layers is constructed using $n-f$ different node previous layer events. Let's denote this property by $P$. 

Take $e$ as any event followed by at least $2f + 1$ different node events. Let $R$ me the maximal base layer followed by any of these events. We state that any layer $L$ greater than $R+1$ and having $P$ property follows $e$. Indeed, let $w$ be any event from $L$. By $P$ property $w$ follows $n-f$ elements from layer $R+1$. Since $n-f + 2f + 1 > n + f$ and $f$ is the maximal possible number of malicious nodes, there's an honest node events $e_1, e_2$ with $e_1$ following $e$ and not following an $R+1$ layer event, and $w$ following $e_2$ from layer $R+1$. Since their node is honest, either $e_1$ follows $e_2$ or vice-versa. First case is impossible, hence $e_2$ follows $e_1$ and $w$ follows $e$. Thus $L$ follows $e$.

Let's take any such layer $L$ following $e$. By lemma statement base layer definitions, any higher than $L$ base layer element follows at least one element from $L$ and thus follows $e$. Taking $k_0$ as the base layer index corresponding to $L$ we finish the proof.
$\square$

After giving some base layer examples, we can pass to voting layer ones.

\subsection{Voting Layer Examples} \label{voting_layers}
In this section we restrict ourselves to three families of voting layer examples. In all cases $k$\textsuperscript{th} voting layer is built upon $k$\textsuperscript{th} base layer. We start with s-voting layer induced by ``strongly seen'' relation, similar to the one used in Hashgraph consensus:
\begin{deff}[S-voting layer]
 Given $k$\textsuperscript{th} base layer events, we construct $k$\textsuperscript{th} \textbf{s-voting layer} as follows: node's event belongs to $k$\textsuperscript{th} s-voting layer if it is node's earliest event which strongly sees at least $n-f$ events from $k$\textsuperscript{th} base layer.
\end{deff}
The above definition inductively gives rise to the following voting layer family:
\begin{deff}[S(m)-voting layer] \label{def.sm-vote}
 Given $k$\textsuperscript{th} base layer events, we inductively construct $k$\textsuperscript{th} \textbf{s(m)-voting layer} ($m\geq 2$) as follows: 
 \begin{itemize}
  \item s(1)-voting layer is equal to s-voting layer from the previous definition
  \item node's event belongs to $k$\textsuperscript{th} s(m)-voting layer if it is node's earliest event that strongly sees at least $n-f$ events from $k$\textsuperscript{th} s(m-1)-voting layer.
 \end{itemize}
\end{deff}
\begin{rem}
 The definition of s(m) voting layers implicitly contains round completion property (cf. definition \ref{round compl}). This slightly differs it from rounds used in Hashgraph consensus. 
\end{rem}
\begin{rem}
 Layer similar to s(2)-voting layer is also mentioned in \cite{Baird:2016:1} as an alternative to s-voting layers.
\end{rem}
Later we will mostly use the following voting layers:
\begin{deff}[S$'$-voting and S$'$(m)-voting layers]\label{def.s'm-vote}
 Replacing ``strongly seeing'' by ``strongly following'' in definitions of s-voting and s(m)-voting layers we get the definitions of \textbf{s$'$-voting} and \textbf{s$'$(m)-voting layers} correspondingly.
\end{deff}
Next voting layer example is naturally connected with a-rounds:
\begin{deff}[A-voting layer]
 Given $k$\textsuperscript{th} base layer events, $k$\textsuperscript{th} \textbf{a-voting layer} is constructed as follows: node's event belongs to $k$\textsuperscript{th} a-voting layer if it is node's earliest event that clearly follows at least $n-f$ events from $k$\textsuperscript{th} base layer.
\end{deff}
The corresponding voting layer family is:
\begin{deff}[A(m)-voting layer] \label{def.am-vote}
 Given $k$\textsuperscript{th} base layer events, $k$\textsuperscript{th} \textbf{a(m)-voting layer} ($m\geq 2$) is inductively constructed as follows: 
 \begin{itemize}
  \item a(1)-voting layer is equal to a-voting layer from the previous definition
  \item node's event belongs to $k$\textsuperscript{th} a(m)-voting layer if it is node's earliest event that clearly follows at least $n-f$ events from $k$\textsuperscript{th} a(m-1)-voting layer.
 \end{itemize}
\end{deff}
Of course, all the above definitions are based just on events and their ancestors. This implies the first condition of being ``appropriate'' from definition \ref{approp_def}. In order to show that the second condition is satisfied we need, first, to provide the fame consensus used.

\subsection{Fame Voting Consensus} \label{fame_consensus}
Here we will present our modified fame voting consensus. Recall that in \textit{decideFame} procedure Hashgraph algorithm uses BFT binary consensuses similar to that of Bracha and Toueg (cf. proposition \ref{hg_bt_sim}). These consensus algorithms have different (though equivalent) decision procedures (cf. end of proposition \ref{hg_bt_sim} proof): 
\begin{itemize}
 \item the former waits until some witness strongly sees more than $2n/3$ of previous round witnesses with the same vote
 \item that of Bracha and Toueg requires that node decides when its newly created witness strongly sees (in relaxed sense, definition \ref{relaxed_ss}) more than $(n + f )/2$ previous round witnesses with the same vote.
\end{itemize}
These decision examples raised a question: ``Is there a faster, yet equivalent decision procedure for a fame consensus?''. An example of such faster procedure is given in the next definition.
\begin{deff}
 We call \textbf{fast decision} (or \textbf{f-decision}, for short) the procedure when a node decides if its newly created event (not necessarily witness) strongly follows more than $(n + f )/2$ same round witnesses with the same vote.
\end{deff}
\begin{lemm}
  F-decision is faster than and equivalent (decides on the same value) to decisions in both Hashgraph fame consensus and Bracha-Toueg one.
\end{lemm}
\textit{Proof.} Consider some node $p$. By definition, if Hashgraph fame consensus decides for $p$, then some witness $w$ (probably created by another node) strongly sees more than $2n/3$ of previous round witnesses with the same vote. Node $p$ is aware of $w$, hence node $p$'s latest event strongly follows all $2n/3$ of previous round witnesses, which are strongly seen by $w$. Since $2n/3 > (n+f)/2$, we got the conditions for f-decision for $p$. This implies that f-decision has weaker requirements than the Hashgraph fame consensus one. Hence it is faster. 

The statement that f-decision is faster than Bracha-Toueg one follows directly from the definition and the fact that ``strongly seeing'' implies ``strongly following''. 

To prove the equivalence of decision procedures, we suppose that f-decision condition is satisfied: $p$'s latest event strongly follows a set of more than $(n+f)/2$ given round witnesses (we denote it by $M_v$) with the same vote $v$, and show that all next round witnesses will have $v$ as a vote (implying that both Hashgraph and Bracha-Toueg fame consensuses will decide $v$). 

The obvious fact that $(n+f)/2 + (n+f)/2 = n + f$ implies that at most one node's witness of the given round can be strongly followed by an event (even for malicious nodes, trying to fork).

Let $w_n$ be any next round witness and $S$ (with $|S|\geq n - f$) be the corresponding set of witnesses strongly seen by it ($|.|$ means the cardinality of a set). Set $S\cap M_v$ contains at least $|M_v| + |S| - n$ elements. Since 
$$|M_v| + |S| - n > \frac{n + f}{2} + |S| - n \geq \frac{n + f}{2} + \frac{n - f}{2} + \frac{|S|}{2} - n = \frac{|S|}{2},$$ 
more than a half elements of $S$ has vote $v$, which implies that $w_n$ will vote $v$. This finishes the proof. $\square$  

With f-decision being defined, we can describe fame consensus procedure that will be used further. 
\begin{deff}[F-fame consensus. Consensus layers] \label{f_fame}
 Suppose that we have an element $e$ from some base layer with $V$ being the corresponding voting layer. We want nodes to collectively decide whether $e$ is famous or not. We define \textbf{f-fame consensus} similarly to that of Hashgraph and Bracha-Toueg fame ones as follows: 
 \begin{itemize}
  \item by analogy with witnesses and base layers, we say that an element belongs to \textbf{1-consensus layer} if it is node's earliest element that strongly follows at least $n-f$ voting layer elements (which we can call also  \textbf{0-consensus layer}, for convenience). 
  \item inductively, \textbf{k-consensus layer} consists of node's earliest elements strongly following at least $n-f$ (k-1)-consensus layer elements
  \item voting layer elements vote for $e$ by whether they clearly follow it or not (and not necessarily seeing it !)
  \item votes are passed to the next consensus layers similarly to Hashgraph and Bracha-Toueg fame consensuses: by majority rule
  \item {\normalfont \textbf{[f-decision]}} whenever we have f-decision condition satisfied: an event strongly follows a set of more than $(n+f)/2$ k-consensus layer elements (for some $k\geq 0$) with the same vote $v$, we decide $v$.
 \end{itemize}
\end{deff}
The proof that f-fame consensus is well-defined is completely analogous to that of Hashgraph and Bracha-Toueg.    

\subsection{Second appropriateness condition proof}
Given examples of base, voting layers, and having defined f-fame consensus, we can prove the following analogue of Baird's lemma 5.17 from \cite{Baird:2016:1}:
\begin{lemm} \label{fame_lemma}
 Let's choose a $k$\textsuperscript{th} g-base layer $B$ (definition \ref{g_base_def}). Then, depending of the choice of $k$\textsuperscript{th} voting layer, under f-fame consensus we have:
 \begin{itemize}
  \item[a.] if voting layer is s(m)-voting or $s'$(m)-voting layer then there will be at least one famous event in $B$ as soon as 2-consensus layer element is created.
  \item[b.] in case of a(m)-voting layer,  as soon as 2-consensus layer element is created, there will be either a famous event in $B$ or a fork $(e, e') \in B$ will be detected with both $e$ and $e'$ followed by voting layer events which, in their turn, are strongly followed by some 1-consensus layer elements.
 \end{itemize}
\end{lemm}
\textit{Proof.} Case (a). The proof in this case is very similar to that of lemma 5.17 from \cite{Baird:2016:1}. We provide it here for convenience. Suppose that 2-consensus layer element $w$ is created. Denote by $S_1$ (with $|S_1| \geq n-f$) the set of 1-consensus events strongly followed by $w$. Let $S_0$ (with $|S_0| \geq n-f$) be the set of voting layer elements strongly followed by at least one element from $S_1$. Each element in $S_0$ strongly follows at least $n-f$ elements from $B$. Since at most one node's event in the given layer can be strongly followed, the fact $n-f > 2n/3$ implies that there's an element $b\in B'\subseteq B$ that is strongly followed by more than $2n/3 * |S_0|/|B'| \geq 2|S_0|/3$ elements from $S_0$ ($B'$ consists of elements from $B$ strongly followed by any element from $S_0$, hence $|B'| \leq n$). Strongly following implies clear following thus more that two thirds of elements from $S_0$ votes for $b$ being famous. We denote by $S_b$ these elements.

Consider any element $e$ from $S_1$. Let $S_e$ be all events from $S_0$ strongly followed by $e$. Then $|S_e| > 2n/3$ and 
$$|S_b\cap S_e| = |S_b| + |S_e| - |S_b\cup S_e| > 2|S_0|/3 + (|S_e|/2 + n/3) - |S_0| \geq |S_e|/2.$$
Hence the majority of $S_e$ will vote TRUE for $b$ being famous, and thus $e$ will. This implies that all elements in $S_1$ will vote TRUE and $w$ decides that $b$ is famous (f-decision condition is satisfied, since $n-f > (n+f)/2$). 

Case (b). We define $w, S_1, S_0$ the same way we did in (a). Here key difference with (a) is that elements from $S_0$ just clearly follow and not necessarily strongly follow events from $B$. Denote by $B'$ events from $B$ clearly followed by at least one event from $S_0$. Two cases are possible: 
\begin{itemize}
 \item There are no forks in $B'$. This implies that only one given node event from $B'$ is clearly followed by events from $S_0$. Hence we can argue the same way we did in (a) to show the existence of famous event $b\in B'$.
 \item For some node there's a fork $(e, e')$ in $B'$. By definition of $B'$ this means that there are events $s, t \in S_0$ such that $s$ clearly follows $e$ and $t$ clearly follows $e'$. Since both $s$ and $t$ are strongly followed by some events from $S_1$, we proved the lemma statement. $\square$   
\end{itemize}

The lemma allows us the prove the desired second ``appropriateness'' condition of base and voting layer examples given above: 

\begin{propo}[Appropriateness, 2nd] \label{approp_propo}
 For g-base layer and any voting layer example provided above, under f-fame consensus, we have that if an event is known to more than $2f$ nodes then it will be followed by a famous event from some base layer.
\end{propo}
\textit{Proof.} Let $w$ be an event from the proposition statement. We consider two cases: 

\textit{Case one: voting layer is s(m)-voting or $s'$(m)-voting layer}. Then by fourth g-base layer property (cf. definition \ref{g_base_def}) there will be g-base layer $L$, such that any element in $L$ follows $w$. By lemma (\ref{fame_lemma}a) there's eventually a famous element in $L$, hence it follows $w$.

\textit{Case two: voting layer is a(m)-voting layer}. Again, by fourth g-base layer property, there will be g-base layer $L_0$ following $w$. By lemma (\ref{fame_lemma}b) there's either a famous element in $L_0$ or a fork $(e, e') \in L_0$ with both $e, e'$ followed by voting layer events which, in their turn, are strongly followed by some 1-consensus layer elements. The former case brings us the proof. We consider the second, the fork case.  

Since both $e$ and $e'$ are strongly followed and $(n + f)/2 > 2f$, we can apply fourth g-base layer property to them. Take any g-base layer $L_1$ following both $e$ and $e'$ (and following $w$, of course). Let $V_1$ be the voting layer corresponding to $L_1$. Denote by $F$ malicious node that created $e$ and $e'$. Since all events in $L_1$ follow $F$'s fork, no event in $V_1$ will clearly follow any events in $L_1$ created by $F$ (if there are any). De-facto $F$ gets banned from the network. By lemma (\ref{fame_lemma}b) there's either a famous element in $L_1$ or a fork (this time by different from $F$ malicious node). And, again, we either get a proof or get second malicious node. Repeating the above step until we ban all forking malicious nodes we will eventually find a famous event following $w$. $\square$  

\subsection{BVCFameDecide Procedure Pseudocode}
Having proven appropriateness of base, voting layer, fame consensus examples, we present pseudocode of \textit{BVCFameDecide} procedure, which can be considered as a merge of general analogues of \textit{divideRounds} and  \textit{decideFame} procedures. \textit{fastFindOrder} procedure, a fast modification of \textit{findOrder} one will be provided in the next subsection. As in Baird's Hashgraph consensus, these procedures will be called by node one after another right after every gossip receipt:

\begin{algoproc}[H]
  \DontPrintSemicolon
  run two loops in parallel:\;
  \While{TRUE} {
    gossip all known events to a random node
  }
   \While{TRUE} {
    receive a gossip\;
    create a new event\;
    \textbf{call} BVCFameDecide\;
    \textbf{call} fastFindOrder
  }
  \caption{General Base-Voting-Consensus atomic broadcast algorithm}
\end{algoproc}
Given appropriate in the sense of definition \ref{approp_def} base, voting and consensus layers, procedure \textit{BVCFameDecide} consists of their calculation and of fame decisions based on f-fame consensuses from definition \ref{f_fame}. In particular, as base layer one can take any example from subsection \ref{base_layers}, as voting layer one can choose $s(m)$, $s'(m)$ or $a(m)$ voting layers (cf. subsection \ref{voting_layers}),  as consensus layers - the ones from definition \ref{f_fame}. We will need the following:
\begin{deff}[(Un)decided base layer] \label{decided_undecided}
 As we will see, {\normalfont BVCFameDecide} includes f-fame consensuses (cf. definition \ref{f_fame}) on whether base layer elements are famous or not. If some base layer has all their possible elements (even not created yet) with decided fame, we call such base layer \textbf{decided}. Otherwise we call it \textbf{undecided}. 
\end{deff}
Now procedure \textit{BVCFameDecide} can be presented in the following preudocode:

\begin{algoproc}[H]
  \DontPrintSemicolon
  \tcc*[l]{Event e runs through all newly added/created events during the last gossip receipt, ordered using follows relation with earliest going first}
  \If{$e$ should be member of a new base layer}{
      \textbf{create} new base layer and \textbf{add} $e$ to it 
  }
  $\mathbb{B}\leftarrow$ set of all undecided base layers, with earliest going first\;
  \For{base layer $B$ in $\mathbb{B}$} {
    \tcc*[l]{Calculating whether e belongs to a particular base layer}
    \If{$e$ should be a member of $B$}{
      \textbf{add} $e$ to $B$\;
      \If{$e\in B$ was declared unfamous in absentia}{
	$e.famous_B\leftarrow$ FALSE \tcp*[r]{$e\in B$ was created ``too late''}
      }
    }
    \tcc*[l]{Starting f-fame consensus protocol for elements from B}
    $K\leftarrow$ maximal non-negative number (if exists) such that $K$-consensus layer $C_K$ corresponding to $B$ has more than $(n + f)/2$ strongly followed elements\;
    \If{$K$ exists}{
      $S\leftarrow$ the set of strongly followed elements in $C_K$\;
      $v\leftarrow$ the majority vote of elements in $S$ for whether $b$ is famous\;
      $N_v\leftarrow$ number of elements in $S$ with vote $v$\;
      \If(\tcp*[f]{f-decision condition}){$N_v > (n+f)/2$}{
	$b.famous_B\leftarrow v$ \tcp*[r]{deciding fame of b as a member of B}
	\If{all possible elements in $B$ have fame decided}{
	  continue loop for the next base layer in $\mathbb{B}$
	}
      }
    }
    $V\leftarrow$ the voting layer corresponding to $B$\;
    \If{$e$ should be a member of $V$}{
      \textbf{add} $e$ to $V$\;
      \For{event $b$ in $B$}{
        $e.vote^V_{b\in B}\leftarrow$ does $e\in V$ clearly follow $b\in B$? \tcp*[r]{voting}
      }
      \If{some possible member $x$ of $B$ is still not created}{
	$e.vote^V_{x\in B}\leftarrow$ FALSE \tcp*[r]{$e\in V$ votes against $x$'s fame in absentia}
      }
  
    }
    \If(\tcp*[f]{$k>0$}){$e$ should be a member of any k-consensus layer $C$ corresponding to $B$}{
      \textbf{add} $e$ to $C$ \tcp*[r]{creating new layer C if necessary}
      \For(\tcp*[f]{including not created yet too}){undecided event $b$ in $B$}{
	$C_{k-1}\leftarrow$ $(k-1)$-consensus layer corresponding to $B$\;
	$S\leftarrow$ the set of elements from $C_{k-1}$ strongly followed by $e$\;
	$v\leftarrow$ the majority vote of elements in $S$ for whether $b\in B$ is famous (TRUE for a tie)\;
	$e.vote^C_{b\in B}\leftarrow v$ \tcp*[r]{voting in higher layers as element of $C$}
      }
    }
  }
  \caption{BVCFameDecide}
\end{algoproc}
\begin{rem}
 We should note that we used the notations $x\in B$ and $b\in B$ in {\normalfont BVCFameDecide} procedure pseudocode to emphasize the fact that they can be members of several base layers. Votes for such elements and their fame are collected and decided separately \underline{regarding the base layer being considered}.       
\end{rem}
\begin{rem}
 Similarly to the previous remark notation $e.vote^C_*$ means that $e$ can have different votes as a member of different consensus layers $C$. 
\end{rem}

\begin{rem}
 In the above procedure we used several simple facts:
 \begin{itemize}
  \item If an event is just added to hashgraph, it can't be strongly followed for $n>1$.
  \item If f-decision condition is satisfied for $k$-consensus layer, it will be also satisfied for all consecutive consensus layers as soon as they have enough number of strongly followed elements.
 \end{itemize}
\end{rem}
\begin{rem}
 As one can see, the decision conditions are checked before doing most of calculations for $e$. This is done for performance reasons. 
\end{rem}

\subsection{Total Order} 
The goal of this subsection is to present a simple modification of \textit{findOrder} procedure with the emphasis on speed and low latency. Informally, by atomic broadcast algorithm \textit{commit latency} we mean average time between event creation and the time when this event is added to the event total order (cf. section \ref{speed_comparison} for formal definition). 
\begin{rem}
Further on we will identify words ``faster'', ``higher speed'' with ``lower latency'' for convenience.    
\end{rem}
As we already mentioned in network model description (cf. \ref{net_model}), consensus protocol is possible with probabilistic termination only. Recall that, to achieve this for fame consensuses, Bracha and Toueg assume in \cite{Bracha:1985:ACB:4221.214134} fair scheduler property, while Bracha in \cite{Bracha:1987:ABA:36888.36891} and Baird in \cite{Baird:2016:1} add randomization using coin toss/flip tricks. In this paper we follow Bracha and Toueg for algorithm definition simplicity, but coin tosses can easily be added if necessary.

Now, when we have fame consensus probabilistic termination we present our \textit{findOrder} procedure version as a pseudocode, starting with the following definitions:
\begin{deff}[Layer processed]
 Recall from definition \ref{decided_undecided} that base layers can be decided or undecided. If we process decided base layer in {\normalfont fastFindOrder} procedure, we set their \textbf{processed} property to TRUE.
\end{deff}
\begin{algoproc}[H]  
  \DontPrintSemicolon
  $\mathbb{B}\leftarrow$ set of all decided and not yet processed base layers\;
  $B_{min}\leftarrow$ the earliest element in $\mathbb{B}$\;
  \While{there is no unprocessed base layers before $B_{min}$}{
    \uIf{there's no famous element in $B_{min}$}
      {\tcc*[l]{Skipping this layer, fork was detected}}
    \Else{
    \tcc*[l]{Any newly added/created event has `commitLayer' and `commitSubLayer' parameters equal to null}  
    $E\leftarrow$ set of all events with null commitLayer and followed by at least one famous event from $B_{min}$ \;
    $subLayer\leftarrow 0$\;
    \While{$E$ is not empty}{
      $E_{min}\leftarrow$ subset of $E$ consisting of events that follow only events with commitLayer not null\;
      \ForAll{$e\in E_{min}$}{
        $e.commitLayer\leftarrow$ ordinal layer number of $B_{min}$\;
        $e.commitSubLayer\leftarrow subLayer$\;
        $e.consensusTimestamp\leftarrow$ median of creation timestamps of all famous events in $B_{min}$        
      }
      $E\leftarrow E \setminus E_{min}$ \tcp*[r]{set subtraction}
      $subLayer\leftarrow subLayer + 1$
    }
    }
    $B_{min}.processed\leftarrow TRUE$ \tcp*[r]{recording the fact being processed}
    $\mathbb{B}\leftarrow\mathbb{B} \setminus\{B_{min}\}$\;
    $B_{min}\leftarrow$ the earliest element in $\mathbb{B}$ \textbf{if} there's any \textbf{else} break the loop
  }
  \tcc*[l]{Similarly to \cite{Baird:2016:1} event's whitened signature is its signature XORed with the signatures of all famous events in event's commit layer $B_{min}$}
  \Return{all events with defined $commitLayer$, sorted by $commitLayer$, then ties sorted by $commitSubLayer$, then by whitened signatures}
  \caption{fastFindOrder}
\end{algoproc}
\begin{rem}
 The main difference from Baird's {\normalfont findOrder} procedure is that we don't wait until event is followed by {\normalfont all} famous layer events. We need just one famous event. Of course, {\normalfont consensusTimestamp} is no more a differentiator for events with the given {\normalfont commitLayer}, but we think this is a good trade-off for the algorithm speed.
\end{rem}
\begin{rem}
 The {\normalfont commitSubLayer} assignment trick is similar to Moser and Melliar-Smith notion of {\normalfont candidate sets} in \cite{Moser:1999:BTO:300547.300555}.
\end{rem}
In order to do the next remark we need the following definition:
\begin{deff}[Supported events] \label{supp_def}
 We call an event \textbf{supported} if it is clearly followed by more than (n + f)/2 events created by different nodes. 
\end{deff}
\begin{rem}
 With enough calculation power a node can do {\normalfont BVCFameDecide} and {\normalfont fastFindOrder} procedures right after every event addition to hashgraph, and not only after gossip receipt during which multiple events usually added. In order for this to take an effect one can replace ``strongly following'' by weaker ``being supported'' property in f-decision part of f-fame consensus definition \ref{f_fame}. This can allow node to faster confirm that certain events were added to the total order. 
\end{rem}
\begin{rem}
 Definition \ref{approp_def}, consensus probabilistic termination and {\normalfont fastFindOrder} procedure guarantee that, given appropriate base/voting layers and consensus definitions, every event known by more than $2f$ nodes will be committed (added to the total order) with probability $1$. They also assure that this order is consistent across the honest nodes. Proposition \ref{approp_propo} assures that the same is true for g-base layers, voting layer examples and f-fame consensus provided in the previous subsections. 
\end{rem}
Having defined BVC atomic broadcast family of algorithms and having shown its richness and appropriateness, we can pass to the next section, concerning one of the most important features of atomic broadcast algorithms: commit latency.

\section{Algorithm Commit Speed Comparison} \label{speed_comparison}
In this section we define algorithm \textit{commit latency} and \textit{commit speed} and show that some BVC family members are almost $1.5$ times faster than Hashgraph consensus algorithm. 

Suitable atomic broadcast speed metrics choice is a hard problem on itself. In theory, one has to know network parameters, adversary possibilities to control the network and possible message exchange scenario distribution. There are many comparative studies of atomic broadcast algorithms some of which can be found, for example, in \cite{DBLP:conf/icccn/UrbanDS00} and \cite{Urban:2004:1}. During such comparisons most widely used metrics are the latency ones, a variant of which called \textit{commit latency} we will use in our comparison:
\begin{deff}[Commit latency] \label{commit_latency}
 \begin{itemize}
  \item Informally, by atomic broadcast algorithm \textbf{commit latency} we mean average time between event creation and the time when this event is added to the event total order.
  \item More formally, let $p$ be a node and $H_p$ its hashgraph representing some gossip exchange scenario. Then for any algorithm $\mathcal{A}$ from BVC family (cf. previous section) we define its \textbf{commit latency} on $H_p$ as the average of differences $t_{comm}^\mathcal{A}(e) - t_{cr}^\mathcal{A}(e)$, where $e\in H_p$ runs through committed by $\mathcal{A}$ events (added to the total order by {\normalfont fastFindOrder} procedure), $t_{cr}^\mathcal{A}(e)$ and $t_{comm}^\mathcal{A}(e)$ are event creation and commit times defined further.      
 \end{itemize}
\end{deff}
For uniformity reasons we don't use event creation timestamp as it's creation time $t_{cr}^\mathcal{A}(e)$, which we define it as follows: 
\begin{deff}[Creation time]
 For an event $e$ its \textbf{creation time} $t_{cr}^\mathcal{A}(e)$ is defined as the length of the longest {\normalfont follows relation} path in hashgraph from event to any node starting event. 
 We assume that hashgraph edges between event and its self-parent have length $0$, while between event and its other parent have length $1$.
\end{deff}
\begin{rem}
 De-facto event creation time measures time passed from network start, with assumption that one gossip takes some fixed amount of time (which we call simply \textbf{unit time}). 
\end{rem}
\begin{deff}[Commit time]
 Let $e$ be an event in honest node $p$ hashgraph $H_p$. Then $e$'s \textbf{commit time} is defined as creation time of earliest created by $p$ event $e'$ whose set of ancestors (which is itself a hashgraph) is enough for algorithm $\mathcal{A}$ to decide that $e$ can be committed.
\end{deff}
An obvious weakness of commit time metric is that if there are few or none events in hashgraph ready to commit then commit time will be inadequate. But it's not really a problem since in real applications it makes sense to consider long enough gossip scenarios whose corresponding hashgraphs will be very big and contain lots of committed events.  

In definition \ref{commit_latency} we defined commit latency for some hashgraph $H_p$. In order to make this definition more objective we consider a representative set of $180$ different gossip scenarios: $20$ scenarios for each node number $n\in\{4, 5, 6, 10, 12, 15, 20, 30, 50\}$ of which $10$ scenarios contain no faults and $10$ contain crash faults whose number is uniformly increasing from $1$ up to $f = \lfloor(n - 1)/3\rfloor$. We extend algorithm \textbf{commit latency} definition on the corresponding set of hashgraphs taking the average of commit latencies on each of them.  

\begin{deff}[Commit speed]
 By BVC algorithm \textbf{commit speed} we simply mean the inverse of its commit latency. With this definition we can freely use words ``faster'', ``slower'', etc.
\end{deff}

Now we describe how the exchange scenarios above were constructed. As the basis we choose computational network model from \cite{Fischer:1985:IDC:3149.214121}. It consists of processes communicating by sending each other messages. A \textit{message} is a pair $(p, m)$, where $p$ is the name of the destination process and $m$ is a ``message value'' from a fixed universe $M$. The message system maintains a multiset, called the \textit{message buffer}, of messages that have been sent but not yet delivered. It supports two abstract operations:
\begin{itemize}
 \item \textit{send(p, m)}: Places $(p, m)$ in the message buffer.
 \item \textit{receive(p)}: Deletes some message $(p, m)$ from the buffer and returns $m$, in which case we say $(p, m)$ is delivered, or returns the special null marker $\varnothing$ and leaves the buffer unchanged.
\end{itemize}
The message system acts nondeterministically, subject only to the condition that if $receive(p)$ is performed infinitely many times, then every message $(p, m)$ in the message buffer is \textit{eventually delivered}.

We apply the above setting to our model, replacing message $(p, m)$ by gossip and its destination pair. 
\begin{rem}
One can note that the ``eventually delivered'' condition above prohibits any message loss, which is not realistic when the network is the Internet itself (cf. an interesting study by Bakr and Keidar in \cite{Bakr:2002:ERT:571825.571864} showing that message loss can sometimes raise above 40\%). In our case message loss will be modeled when node's earlier gossip will be delayed and stay longer in message buffer than its more recent gossip. It can happen that the former gossip will bring no new knowledge to its destination node and hence will be skipped by it and considered as being lost.     
\end{rem}

We construct gossip scenarios for $n$ nodes with $k$ crash-faulty nodes as follows:
\begin{itemize}
 \item each node has one starting event
 \item message buffer total operation number is set to $1000 \cdot n$ 
 \item on each step we randomly choose which message buffer operation to choose: $send$ or $receive$, each with probability $0.5$
 \item $k$ faulty nodes are randomly chosen, as well as their crash fault times 
 \item when $send$ is chosen, we randomly choose non-crashed node $q$ and non-crashed destination node $q$ and place gossip from $p$ to $q$ in the message buffer
 \item when $receive$ is chosen, we randomly take any gossip from the message buffer
\end{itemize}

Finally, the corresponding hashgraphs being considered are those of the node with $id = 0$ after all $1000\cdot n$  message buffer operations being completed.
\begin{rem}
 For more clarity we put the *.csv files of all $180$ generated scenarios on \url{https://github.com/trafim/gossip_scenarios}. Interested readers can conduct all calculations on them themselves.
\end{rem}
\begin{notei}[BVC examples]
In order to do the comparison we will use simple notations for BVC family members given in section \ref{section.BVC}. Namely, we denote a BVC algorithm by $\mathbf{BVC.\langle base\ layer\rangle.\langle voting\ layer\rangle}$ where:
\begin{itemize}
 \item $\langle$base layer$\rangle$  is $\mathbf{A}$ for a-base layer (definition \ref{def.a-base}), $\mathbf{S}$ for s-base layer (definition \ref{def.s-base}), $\mathbf{S'}$ for $s'$-base layer (definition \ref{def.s'-base}), $\mathbf{C_{a,b}}$ and $\mathbf{C'_{a,b}}$ for base layers $C(a,b)$, $C'(a,b)$ from definitions \ref{def.c-base} and \ref{def.c'-base} correspondingly.   
 \item $\langle$voting layer$\rangle$ is $\mathbf{A_m}$ for a(m)-voting layer (definition \ref{def.am-vote}), $\mathbf{S_m}$ for s(m)-voting layer (definition \ref{def.sm-vote}) and $\mathbf{S'_m}$ for $s'$(m)-voting layer (definition \ref{def.s'm-vote}). 
 \end{itemize}
\end{notei}
\begin{rem}
 We do not include consensus layer in our notation since the same f-fame consensus is used in all BVC examples considered.
\end{rem}
Using the above notation we introduce the following
\begin{notei}
 For algorithm $\mathbf{BVC.S.S_1}$, which is the closest to Hashgraph consensus BVC family member, we will use special $\mathbf{BVC.HG}$ notation. Hashgraph consensus itself will be denoted simply by $\mathbf{HG}$.
\end{notei}
Now we are ready to present a list of algorithms, which will be used in our comparison. They will be: 
\begin{itemize}
 \item $\mathbf{HG}$, $\mathbf{BVC.HG}$, $\mathbf{BVC.S'.S'_1}$
 \item $\mathbf{BVC.A.A_1}$, $\mathbf{BVC.A.A_2}$, $\mathbf{BVC.A.S'_1}$,
 \item $\mathbf{BVC.S'.A_1}$, $\mathbf{BVC.S'.S'_2}$,
 \item $\mathbf{BVC.C_{2, M}.A_1}$ (here and later $M = 10000$), $\mathbf{BVC.C_{2, M}.S'_1}$,
 \item $\mathbf{BVC.C'_{1, M}.A_1}$, $\mathbf{BVC.C'_{1, M}.S'_1}$, $\mathbf{BVC.C'_{2, M}.A_1}$, 
 \item $\mathbf{BVC.C'_{2, M}.S'_1}$, $\mathbf{BVC.C'_{3, M}.S'_1}$, $\mathbf{BVC.C'_{4, M}.S'_1}$, $\mathbf{BVC.C'_{5, M}.S'_1}$
\end{itemize}
We consider the above list of algorithms to be representative enough. We prefer $s'$ layers rather than $s$ layers since they make the algorithms to be a little faster (as we will see). One of the reasons for that is that in $s'$ case an event can be a member of several consensus layers and sometimes helps to make fame decisions earlier. Constant $M$ is taken big enough to measure the impact of $C(a, *)$- and $C'(a, *)$-base layers without considering rare divisible by $M$ layers.

We present commit latency of each of these algorithms in Table \ref{tbl:latency_stats}. We calculate commit latency on all scenarios and also separately on all $20$ scenarios for each  $n\in\{4, 5, 6, 10, 12, 15, 20, 30, 50\}$. Table \ref{tbl:comp_latency_stats} represents comparative commit latencies of these algorithms to that of $\mathbf{BVC.C'_{3, M}.S'_1}$, which is among the fastest chosen algorithms.

We also present commit latency of some of these algorithms in a more convenient way on Figure \ref{plt:latency_stats}. We chose only the most representative algorithms to not overburden the image.  

As one can see Hashgraph consensus is $\mathbf{1.47}$ times slower than the fastest $\mathbf{BVC.C'_{3, M}.S'_1}$. At the same time Hashgraph consensus closest BVC family member $\mathbf{BVC.HG}$ appears to be fairly faster than it, and is $\mathbf{1.16}$ times slower than $\mathbf{BVC.C'_{3, M}.S'_1}$.

\begin{figure}
 \begin{minipage}{.32\linewidth}
  \centering
  \subfloat[]{\label{fig:commit_comp_a}\includegraphics[scale=.6]{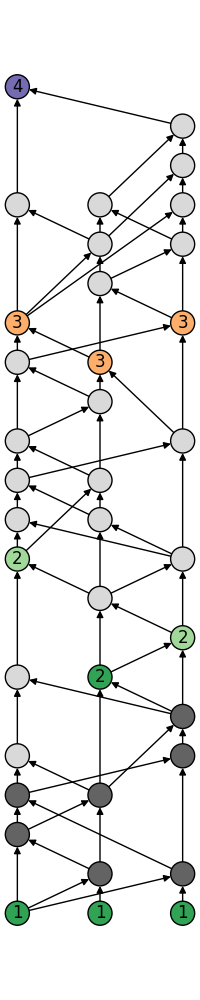}}
 \end{minipage}
 \begin{minipage}{.32\linewidth}
  \centering
  \subfloat[]{\label{fig:commit_comp_b}\includegraphics[scale=.6]{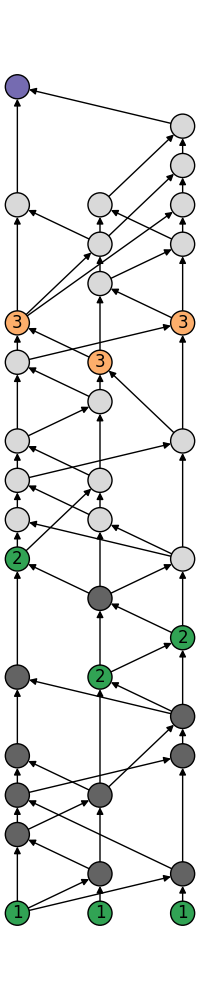}} 
 \end{minipage}
 \begin{minipage}{.32\linewidth}
  \centering
  \subfloat[]{\label{fig:commit_comp_c}\includegraphics[scale=.6]{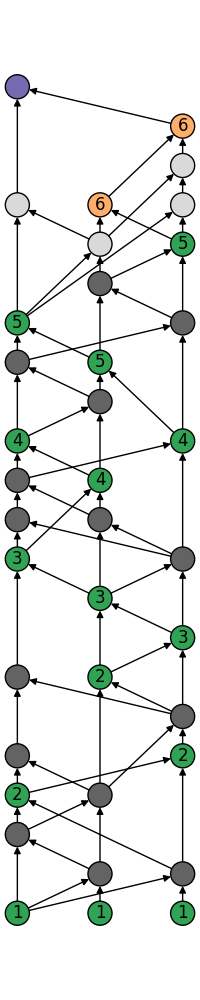}}
 \end{minipage}\par\medskip
 \caption{Figures (\ref{fig:commit_comp_a}), (\ref{fig:commit_comp_b}) and (\ref{fig:commit_comp_c}) represent witnesses (base layer elements) and committed events by algorithms $\mathbf{HG}$, $\mathbf{BVC.HG}$ and $\mathbf{BVC.S'.S'_1}$ correspondingly. Here we use the color notation similar to that of Hashgraph consensus examples \cite{Baird:2016:2}: light orange are witnesses with undecided fame; light green - famous witnesses, light gray - other events; darker colors mean that an event was committed; last event is marked with purple color for convenience. One can see that with the same knowledge $\mathbf{BVC.HG}$ commits more events than $\mathbf{HG}$, and $\mathbf{BVC.S'.S'_1}$ even more than $\mathbf{BVC.HG}$. }
 \label{fig:commit_comp}
\end{figure}

General conclusions are that $s'$-voting layer is more efficient than $a$-, $a(2)$- or $s'(2)$-voting layers. Also that algorithms of type $\mathbf{BVC.C'_{K, M}.S'_1}$ are usually faster. What is the optimal parameter $K$ or how it depends on $n$ and $M$ is an open question. 
\begin{rem}
Algorithm $\mathbf{BVC.A.S'_1}$ is also worth consideration since it is only $1.04$ times slower and at the same time is free from any parameters.   
\end{rem}
\begin{rem}
 To illustrate differences in commit latency we show on Figure \ref{fig:commit_comp} how algorithms $\mathbf{HG}$, $\mathbf{BVC.HG}$ and $\mathbf{BVC.S'.S'_1}$ commit events for randomly generated scenario with $3$ nodes.
\end{rem}

\begin{figure}
\begin{tikzpicture}
  \begin{axis}[
    height=15cm,
    width=15cm,
    grid=major,
    xlabel=Node number,
    ylabel=Latency (in unit time),
    legend pos=south east,
    legend cell align=left
  ]
  \addplot coordinates {(4, 12.9) (5, 17.5) (6, 21.5) (10, 25.7) (12, 31.0) (15, 34.6) (20, 39.6) (30, 45.6) (50, 55.0) };
  \addlegendentry{$\mathbf{HG}$}
  \addplot coordinates {(4, 10.3) (5, 14.2) (6, 15.7) (10, 20.5) (12, 23.8) (15, 26.8) (20, 31.7) (30, 36.0) (50, 43.9) };
  \addlegendentry{$\mathbf{BVC.HG}$}
  \addplot coordinates {(4, 10.1) (5, 14.8) (6, 13.6) (10, 25.5) (12, 26.6) (15, 31.4) (20, 39.9) (30, 45.7) (50, 58.6) };
  \addlegendentry{$\mathbf{BVC.A.A_1}$}
  \addplot coordinates {(4, 9.4) (5, 12.9) (6, 13.0) (10, 18.8) (12, 21.1) (15, 23.8) (20, 29.0) (30, 32.5) (50, 39.9) };
  \addlegendentry{$\mathbf{BVC.A.S'_1}$}
  \addplot coordinates {(4, 14.2) (5, 19.9) (6, 19.8) (10, 28.2) (12, 32.1) (15, 36.4) (20, 43.9) (30, 48.8) (50, 60.1) };
  \addlegendentry{$\mathbf{BVC.S'.S'_2}$}
  \addplot coordinates {(4, 9.5) (5, 12.6) (6, 12.2) (10, 18.2) (12, 20.0) (15, 22.9) (20, 27.9) (30, 30.9) (50, 38.7) };
  \addlegendentry{$\mathbf{BVC.C'_{3, M}.S'_1}$}
  \end{axis}
\end{tikzpicture}
\caption{Commit latency for chosen atomic broadcast algorithms and node numbers.}
  \label{plt:latency_stats}
\end{figure}
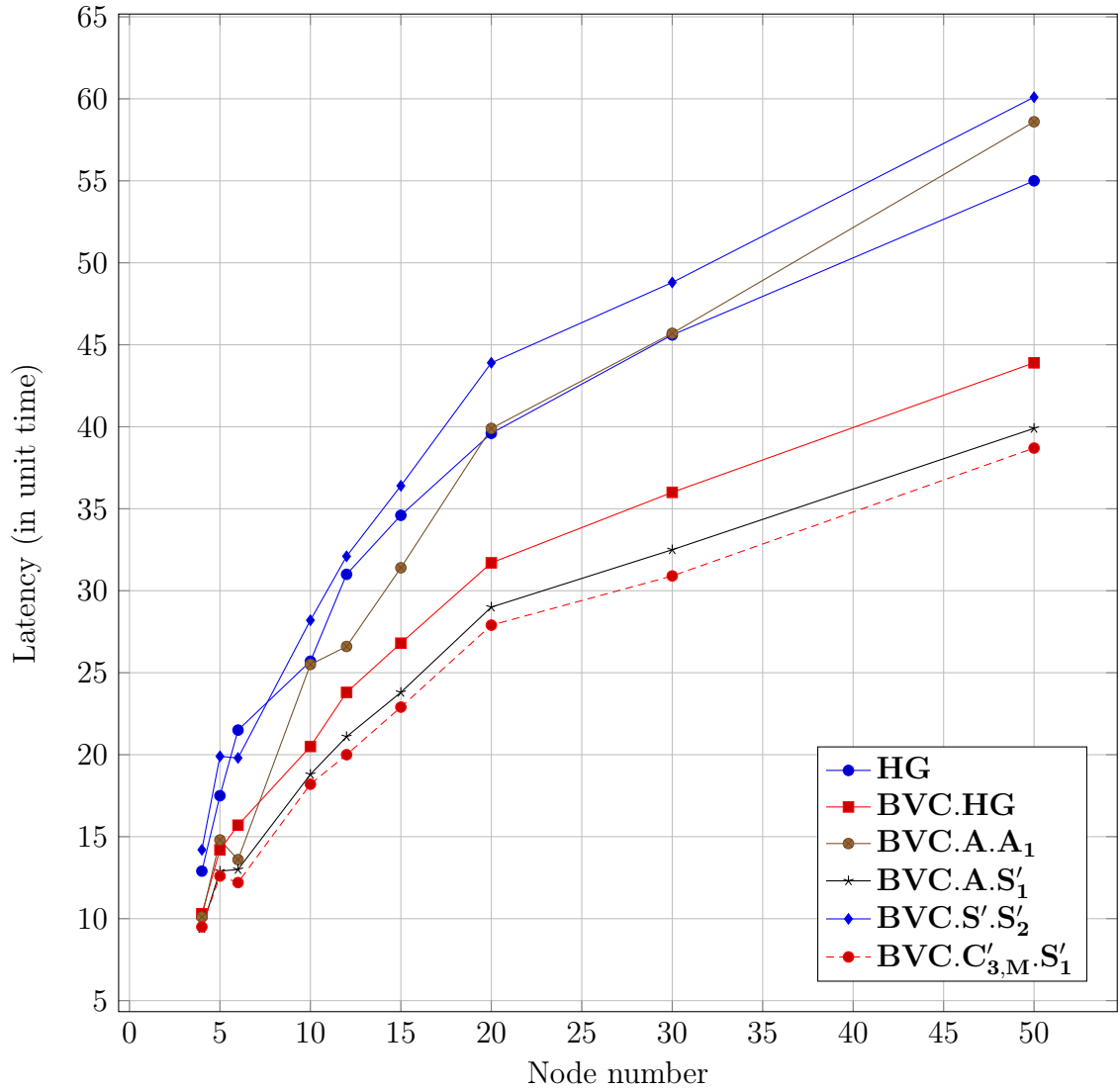
 
\begin{figure}
  \centering
  \begin{tikzpicture}
    \clip node (m) [matrix,matrix of nodes, fill=black!20,inner sep=0pt, nodes in empty cells,
    nodes={minimum height=1cm,minimum width=1.24cm,anchor=center,outer sep=0,font=\sffamily},
    row 1/.style={nodes={fill=black,text=white}},
    column 1/.style={nodes={fill=gray,text=white,align=right,text width=3cm,text depth=0.5ex}},
    column 2/.style={text width=1.1cm,align=center,every even row/.style={nodes={fill=white}}},
    column 3/.style={text width=1.1cm,align=center,every even row/.style={nodes={fill=white}}},
    column 4/.style={text width=1.1cm,align=center,every even row/.style={nodes={fill=white}}},
    column 5/.style={text width=1.1cm,align=center,every even row/.style={nodes={fill=white}}},
    column 6/.style={text width=1.1cm,align=center,every even row/.style={nodes={fill=white}}},
    column 7/.style={text width=1.1cm,align=center,every even row/.style={nodes={fill=white}}},
    column 8/.style={text width=1.1cm,align=center,every even row/.style={nodes={fill=white}}},
    column 9/.style={text width=1.1cm,align=center,every even row/.style={nodes={fill=white}}},
    column 10/.style={text width=1.1cm,align=center,every even row/.style={nodes={fill=white}}},
    column 11/.style={text width=1.1cm,align=center,every even row/.style={nodes={fill=white}}},
    row 1 column 1/.style={nodes={fill=gray}},
    prefix after command={[rounded corners=4mm] (m.north east) rectangle (m.south west)}] 
    {
        & n = 4 & n = 5 & n = 6 & n = 10 & n = 12 & n = 15 & n = 20 & n = 30 & n = 50 & Total\\
      $\mathbf{HG}$ & \textbf{12.9} & \textbf{17.5} & \textbf{21.5} & \textbf{25.7} & \textbf{31.0} & \textbf{34.6} & \textbf{39.6} & \textbf{45.6} & \textbf{55.0} & \textbf{31.5} \\
      $\mathbf{BVC.HG}$ & 10.3 & 14.2 & 15.7 & 20.5 & 23.8 & 26.8 & 31.7 & 36.0 & 43.9 & 24.8 \\
      $\mathbf{BVC.S'.S'_1}$ & 10.4 & 14.2 & 14.1 & 20.5 & 23.0 & 26.3 & 31.7 & 35.5 & 43.9 & 24.4 \\
      $\mathbf{BVC.A.A_1}$ & 10.1 & 14.8 & 13.6 & 25.5 & 26.6 & 31.4 & 39.9 & 45.7 & 58.6 & 29.6 \\
      $\mathbf{BVC.A.A_2}$ & 9.5 & 12.9 & 13.9 & 18.9 & 21.7 & 24.3 & 29.1 & 32.5 & 40.0 & 22.5 \\
      $\mathbf{BVC.A.S'_1}$ & 9.4 & 12.9 & 13.0 & 18.8 & 21.1 & 23.8 & 29.0 & 32.5 & 39.9 & 22.3 \\
      $\mathbf{BVC.S.A_1}$ & 10.8 & 15.6 & 14.7 & 26.3 & 28.1 & 33.3 & 41.5 & 47.6 & 60.6 & 30.9 \\
      $\mathbf{BVC.S'.A_1}$ & 10.8 & 15.6 & 14.4 & 26.3 & 27.9 & 33.1 & 41.5 & 46.9 & 60.6 & 30.8 \\
      $\mathbf{BVC.S'.S'_2}$ & 14.2 & 19.9 & 19.8 & 28.2 & 32.1 & 36.4 & 43.9 & 48.8 & 60.1 & 33.7 \\
      $\mathbf{BVC.C_{2, M}.A_1}$ & 10.4 & 15.0 & 13.6 & 26.7 & 28.1 & 33.9 & 43.5 & 50.1 & 65.3 & 31.9 \\
      $\mathbf{BVC.C_{2, M}.S'_1}$ & 9.2 & 12.2 & 12.0 & 18.1 & 19.8 & 22.9 & 28.2 & 31.5 & 39.7 & 21.5 \\
      $\mathbf{BVC.C'_{1, M}.A_1}$ & 10.5 & 15.1 & 14.0 & 28.2 & 29.2 & 35.4 & 45.1 & 53.1 & 68.5 & 33.2 \\
      $\mathbf{BVC.C'_{1, M}.S'_1}$ & 9.1 & 12.1 & 11.9 & 18.1 & 19.9 & 23.0 & 28.8 & 31.7 & 40.2 & 21.6 \\
      $\mathbf{BVC.C'_{2, M}.A_1}$ & 10.2 & 14.8 & 13.5 & 26.5 & 27.7 & 33.2 & 42.8 & 49.2 & 64.0 & 31.3 \\
      $\mathbf{BVC.C'_{2, M}.S'_1}$ & 9.2 & 12.3 & 12.1 & 18.1 & 19.8 & 23.0 & 28.1 & 31.0 & 39.2 & 21.4 \\
      $\mathbf{BVC.C'_{3, M}.S'_1}$ & \textbf{9.5} & \textbf{12.6} & \textbf{12.2} & \textbf{18.2} & \textbf{20.0} & \textbf{22.9} & \textbf{27.9} & \textbf{30.9} & \textbf{38.7} & \textbf{21.4} \\
      $\mathbf{BVC.C'_{4, M}.S'_1}$ & 9.5 & 12.9 & 12.5 & 18.2 & 20.1 & 22.8 & 28.0 & 31.1 & 38.8 & 21.5 \\
      $\mathbf{BVC.C'_{5, M}.S'_1}$ & 9.5 & 12.9 & 13.0 & 18.4 & 20.3 & 22.9 & 28.1 & 31.1 & 38.8 & 21.7 \\
    };
  \end{tikzpicture}
  \captionof{table}{Commit latency for chosen atomic broadcast algorithms and node numbers.}
  \label{tbl:latency_stats}
\end{figure}

\begin{figure}
  \centering
  \begin{tikzpicture}
    \clip node (m) [matrix,matrix of nodes, fill=black!20,inner sep=0pt, nodes in empty cells,
    nodes={minimum height=1cm,minimum width=1.24cm,anchor=center,outer sep=0,font=\sffamily},
    row 1/.style={nodes={fill=black,text=white}},
    column 1/.style={nodes={fill=gray,text=white,align=right,text width=3cm,text depth=0.5ex}},
    column 2/.style={text width=1.1cm,align=center,every even row/.style={nodes={fill=white}}},
    column 3/.style={text width=1.1cm,align=center,every even row/.style={nodes={fill=white}}},
    column 4/.style={text width=1.1cm,align=center,every even row/.style={nodes={fill=white}}},
    column 5/.style={text width=1.1cm,align=center,every even row/.style={nodes={fill=white}}},
    column 6/.style={text width=1.1cm,align=center,every even row/.style={nodes={fill=white}}},
    column 7/.style={text width=1.1cm,align=center,every even row/.style={nodes={fill=white}}},
    column 8/.style={text width=1.1cm,align=center,every even row/.style={nodes={fill=white}}},
    column 9/.style={text width=1.1cm,align=center,every even row/.style={nodes={fill=white}}},
    column 10/.style={text width=1.1cm,align=center,every even row/.style={nodes={fill=white}}},
    column 11/.style={text width=1.1cm,align=center,every even row/.style={nodes={fill=white}}},
    row 1 column 1/.style={nodes={fill=gray}},
    prefix after command={[rounded corners=4mm] (m.north east) rectangle (m.south west)}] 
    {
        & n = 4 & n = 5 & n = 6 & n = 10 & n = 12 & n = 15 & n = 20 & n = 30 & n = 50 & Total\\
      $\mathbf{HG}$ & \textbf{1.35} & \textbf{1.38} & \textbf{1.75} & \textbf{1.41} & \textbf{1.55} & \textbf{1.51} & \textbf{1.42} & \textbf{1.47} & \textbf{1.42} & \textbf{1.47} \\
      $\mathbf{BVC.HG}$ & 1.08 & 1.13 & 1.29 & 1.13 & 1.19 & 1.17 & 1.14 & 1.16 & 1.13 & 1.16 \\
      $\mathbf{BVC.S'.S'_1}$ & 1.09 & 1.13 & 1.15 & 1.13 & 1.15 & 1.15 & 1.14 & 1.15 & 1.13 & 1.14 \\
      $\mathbf{BVC.A.A_1}$ & 1.06 & 1.17 & 1.11 & 1.4 & 1.33 & 1.37 & 1.43 & 1.48 & 1.52 & 1.38 \\
      $\mathbf{BVC.A.A_2}$ & 0.99 & 1.02 & 1.14 & 1.04 & 1.08 & 1.06 & 1.04 & 1.05 & 1.04 & 1.05 \\
      $\mathbf{BVC.A.S'_1}$ & 0.99 & 1.02 & 1.06 & 1.03 & 1.05 & 1.04 & 1.04 & 1.05 & 1.03 & 1.04 \\
      $\mathbf{BVC.S.A_1}$ & 1.13 & 1.24 & 1.2 & 1.45 & 1.4 & 1.45 & 1.49 & 1.54 & 1.57 & 1.44 \\
      $\mathbf{BVC.S'.A_1}$ & 1.13 & 1.24 & 1.17 & 1.45 & 1.4 & 1.44 & 1.49 & 1.52 & 1.57 & 1.44 \\
      $\mathbf{BVC.S'.S'_2}$ & 1.49 & 1.58 & 1.62 & 1.55 & 1.61 & 1.59 & 1.57 & 1.58 & 1.55 & 1.57 \\
      $\mathbf{BVC.C_{2, M}.A_1}$ & 1.09 & 1.19 & 1.11 & 1.47 & 1.41 & 1.48 & 1.56 & 1.62 & 1.69 & 1.49 \\
      $\mathbf{BVC.C_{2, M}.S'_1}$ & 0.96 & 0.97 & 0.98 & 1.0 & 0.99 & 1.0 & 1.01 & 1.02 & 1.03 & 1.0 \\
      $\mathbf{BVC.C'_{1, M}.A_1}$ & 1.1 & 1.2 & 1.14 & 1.55 & 1.46 & 1.54 & 1.62 & 1.72 & 1.77 & 1.55 \\
      $\mathbf{BVC.C'_{1, M}.S'_1}$ & 0.96 & 0.96 & 0.97 & 0.99 & 1.0 & 1.0 & 1.03 & 1.03 & 1.04 & 1.01 \\
      $\mathbf{BVC.C'_{2, M}.A_1}$ & 1.07 & 1.17 & 1.1 & 1.46 & 1.39 & 1.45 & 1.53 & 1.59 & 1.65 & 1.46 \\
      $\mathbf{BVC.C'_{2, M}.S'_1}$ & 0.96 & 0.98 & 0.99 & 1.0 & 0.99 & 1.0 & 1.01 & 1.0 & 1.01 & 1.0 \\
      $\mathbf{BVC.C'_{3, M}.S'_1}$ & \textbf{1.0} & \textbf{1.0} & \textbf{1.0} & \textbf{1.0} & \textbf{1.0} & \textbf{1.0} & \textbf{1.0} & \textbf{1.0} & \textbf{1.0} & \textbf{1.0} \\
      $\mathbf{BVC.C'_{4, M}.S'_1}$ & 1.0 & 1.02 & 1.02 & 1.0 & 1.0 & 0.99 & 1.0 & 1.0 & 1.0 & 1.0 \\
      $\mathbf{BVC.C'_{5, M}.S'_1}$ & 1.0 & 1.02 & 1.06 & 1.01 & 1.01 & 1.0 & 1.01 & 1.0 & 1.0 & 1.01 \\
    };
  \end{tikzpicture}
  \captionof{table}{Comparative commit latency for chosen atomic broadcast algorithms and node numbers.}
  \label{tbl:comp_latency_stats}
\end{figure}

\newpage

\section{Practical Use Cases, Remarks, Future Work and Acknowledgement} \label{section.practical_future}
\subsection{Applications}
As atomic broadcast algorithms BVC Family can be used in all corresponding domains. Examples are distributed databases and file storages, synchronization services, cryptocurrency and blockchain domains. 

\subsection{Proof-of-Stake}
A special but important case mentioned in \cite{Baird:2016:1} in the proof-of-stake one. We briefly remind it. The idea is that in Hashgraph consensus when node made some action, it added ``+1'' to certain statistics, be it fame voting, decision making, forking, etc. It is not hard to check that if instead node will add not ``+1'' but some positive integer associated with them, known as their ``stake'', all the proofs in Hashgraph consensus still hold. This is due to the fact that Hashgraph consensus depends on properties like ``more than $2n/3$ of the members'' and ``at least half of events'', etc.    

Stakes allow to make some nodes more important than the others and also introduce economic incentives to keep the network alive. Straightforward examples are: reward nodes increasing their stake for creating famous witnesses and punish them for forking decreasing it.  
\begin{rem}
 The above proof-of-stake case can also be extended to BVC Family since its members are based on similar to Hashgraph consensus statistics. The only change that should be made is to replace $f = \lfloor(n - 1)/3\rfloor$ by $n/3$ since stakes are not integer numbers anymore.
\end{rem}

\subsection{Crash Recovery Models}
Another interesting case is ZooKeeper's atomic broadcast protocol \cite{Medeiros:2012:1} system model. It admits that instead of any malicious behavior nodes can just crash and recover indefinitely many times. One can show that all BVC Family algorithms can be applied to this case even when we increase possible number of faulty nodes $f$ from $\lfloor(n - 1)/3\rfloor$ to $\lfloor(n - 1)/2\rfloor$. Proof-of-Stake case is also valid if we change $f < n/3$ to $f < n/2$.  

\subsection{Future Work}
Despite the fact that we presented some quite fast examples of atomic broadcast algorithms, which can be readily used in practice, the work in not finished yet. As the future research direction we consider choosing the most promising network probability models and finding the fastest algorithms in BVC Family according to them.

\subsection{Acknowledgement}
I would like to thank my friends Dr. Pavel Irzhauski, Dr. Alexandr Usnich and  Dr. Evgeniy Zorin for useful discussions and partial review of this work.

\newpage

\appendix

\begin{appendices}
\section{Variant of Gossip Protocol} \label{our_gossip_proto}
Recall that in \textit{gossip protocol} definition (cf. subsection \ref{main_defs}) node sends to another node ``all of the information it knows so far''. Of course knowledge exchange should be optimized since node's knowledge grows linearly with time. In \cite{Baird:2016:2} Baird proposes the following gossip protocol simple optimization:
\begin{itemize}
 \item before gossip receipt from node $i$, node $j$ sends to $i$ an array of how many events $j$ knows about that were created by each node.
 \item node $i$ compares this array with its array and sends to $j$ only events $i$ knows about but $j$ doesn't. For example if $i$ knows $7$ events created by some node $p$ and $j$ knows $5$ events by $p$ then $i$ will send $j$ only two last events created by $p$. 
\end{itemize}
This approach works until some node $p$ starts to fork. In this case $p$'s events and self-parent relation form a tree and not a sequence anymore. Hence during a gossip just sending the number of known events created by $p$ is not enough for knowledge exchange. In this appendix we present an extension of Baird's optimization that allows honest nodes to efficiently share their knowledge about events, ``follows'' relation and forks. 

First, we should mention that malformed events (the ones with two self-parents, wrong transactions, etc.) are filtered out by honest nodes immediately: an honest node just checks each event it receives during gossip using any prescribed event creation rules. If this check fails faulty node, which created this event, is put in the honest node ban list and no gossips from it are ever admitted. 
For convenience we introduce the following definitions:
\begin{deff}[Event Index]
 Let $i$ be some node that didn't fork. Then for an event $e$ created by $i$ we can define it's \textbf{index} as $e$'s ordinary index in created by $i$ event sequence (with starting event having index $0$). We denote this index by $idx_e$.
\end{deff}
\begin{deff}[Last Index Array]
 We call the array of all non-forking node last event indices known to node $i$ as $i$'s \textbf{last index array} and denote it by $\mathbf{lia}_i$.
\end{deff}
\begin{deff}[Digest]
 For an event $e$ created by node $p$ we call its \textbf{digest} triplet $(h(e), id_p, idx_e)$ where $h(e)$ is hash of $e$, $id_p$ is a numeric identifier of $p$ and $idx_e$ is index of $e$.
\end{deff}
\begin{rem}
 In order to exclude certain ambiguities we suppose that each event contains not just the hashes of its parents but their \underline{digests}.
\end{rem}
Now we start to describe our variant of gossip protocol. We consider any two nodes $i$ and $j$. Suppose node $j$ wants to receive all events node $i$ knows about. We describe four cases depending on whether these nodes are aware about any forks or not. We cover first case in details and briefly the other cases.

\subsection{Case 1: Both $i$ and $j$ are not aware of any forks}
Node $j$ starts by sending $\mathbf{lia}_j$ to node $i$. After getting it from $j$, $i$ compares this array with $\mathbf{lia}_i$ and sends to $j$ only events whose indices are greater than the  last index in $\mathbf{lia}_j$ of their corresponding creator. On receipt of these new events from $i$, $j$ sorts them (with earliest going first) using the ``follows'' relation (by looking at parent digests they contain). Then $j$ adds these events one-by-one to its hashgraph checking each time whether new event $e$ to be added is not malformed . Node $j$ also each time verifies two facts for $e$:
\begin{itemize}
 \item Self-parent digest contained in $e$ should coincide with the digest of event from $j$'s hashgraph created by the same node and having smaller by $1$ index than that of $e$. Of course the latter event should exist in $j$'s hashgraph.
 \item Other parent digest in $e$ should coincide with event already present in $j$'s hashgraph. 
\end{itemize}
If these checks are passed then communication is considered to be successfully finished. As one can see in this case the protocol is exactly Baird's optimization described at the beginning of the appendix. 

If $i$ has sent $j$ its known events according to the protocol (otherwise $j$ would cancel gossip receipt) then in case the above facts are not true for one or several events, node $j$ has two possibilities for any such event $e$ created by some node $p$:
\begin{itemize}
 \item {[Self-parent problem]} Event $e$ contains self-parent digest $(h, id_p, idx_e - 1)$ and $j$'s hashgraph has event $e'$ with digest $(h(e')\neq h, id_p, idx_e - 1)$. Since both $e$, $e'$ are created by the same node $p$ we have that either $e$ is malformed or $p$ created a fork $(e, e')$. In both cases $j$ has a proof that $p$ is malicious since all events in a fork are signed by their cheating creator $p$. Using this proof $j$ can issue any kind of punishing $p$ transaction.
 \item {[Other parent problem]} Event $e$ contains other parent digest $(h, id_{p'}, idx)$, where $p\neq p'$ and $j$ knows about event $e'$ with digest $(h'\neq h, id_{p'}, idx_{e'} = idx)$. Here $j$ suspects $p'$ in creating a fork or $p$ in creating malformed event $e$ and $i$ sending $j$ this malformed event nevertheless.
\end{itemize}
In both cases $j$ will not consider communication finished and it will send to $i$ a special case of so-called ``compressed fork information'', which we define in general as follows:

\begin{deff}
 Let $p_f$ be a faulty node who created one or several forks. Then events created by $p_f$ and ``follows'' relation form not a sequence but a tree. Any honest node $p_h$ knows whole or a part of this tree. Node $p_h$ represents its knowledge about $p_f$'s event tree as an array of tuples of hashes $(h(s_0), h(t_0)), \ldots, (h(s_m), h(t_m))$ where $t_0,\ldots, t_m$ are branch tips of $p_f$'s tree known to $p_h$ and $s_i$ represent the latest supported event created by $p_f$ and followed by $t_i$ in $p_h$'s hashgraph. 
 \begin{itemize}
  \item We call this array of tuples $p_h$'s \textbf{compressed fork information} about $p_f$'s event tree. We denote it $\mathbf{cfi}_{p_f}$.
  \item For any tuple $(h(s), h(t))$ from above we define as corresponding \textbf{tree branch} the chain of events $s,\ldots, t$ where each event is the self-parent of the next one.  
 \end{itemize}
\end{deff}
\begin{rem}
 $\mathbf{cfi}_{p}$ can be considered as $p$'s event number replacement for malicious $p$ during gossip exchange between nodes.
\end{rem}
\begin{rem}[Supported]
 We remind from definition \ref{supp_def} that an event is \textbf{supported} if it is clearly followed by more than $(n + f)/2$ events created by different nodes. We use such events because of their nice property that supported events created by the same node (even malicious one) can never form a fork. 
\end{rem}
\begin{rem} \label{more_cfi}
 Compressed fork information can be transferred in an even more compact form when tuples with common $s_i$ are joined together in an array-hash tuple. We provide an example of this on Figure \ref{fig:fork2}. 
\end{rem}
Now we return to our situation when node $j$ just detected one or several forks or suspects a fork or malformed event creation. Since both $i$ and $j$ were not aware of any forks before this communication, to get all events still unknown to $j$ it suffices for $j$ to send $i$ only one tuple $(h(s), h(t))$ for every cheating node $p$ (here $t$ is the last $p$'s event known to $j$ and $s$ it's latest supported ancestor). Note that these tuples represent a particular case of \textit{compressed fork information} received by $i$. We know that supported events form a sequence even for a forking node. This fact allows $i$ to derive which events $j$ is still missing and send them to it. In its turn $j$ figures out that either $i$ was cheating and create punishing $i$ transaction or get a proof of fork. In this case, again, using this proof $j$ can issue any kind of punishing forking node transaction.    

\begin{figure}
 \begin{minipage}{.32\linewidth}
  \centering
  \subfloat[]{\label{fig:fork1a}\includegraphics[scale=.7]{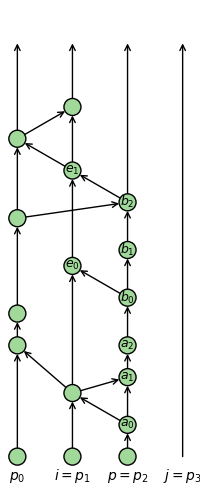}}
 \end{minipage}
 \begin{minipage}{.32\linewidth}
  \centering
  \subfloat[]{\label{fig:fork1b}\includegraphics[scale=.7]{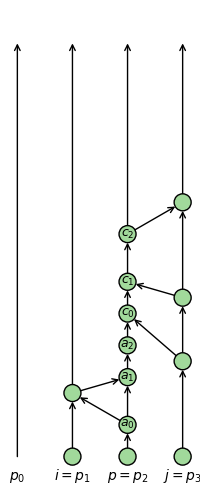}} 
 \end{minipage}
 \begin{minipage}{.32\linewidth}
  \centering
  \subfloat[]{\label{fig:fork1c}\includegraphics[scale=.7]{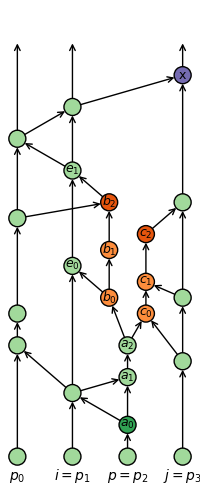}}
 \end{minipage}\par\medskip
 \caption{Case 1 of gossip protocol variant. (\ref{fig:fork1a}), (\ref{fig:fork1b}) represent hashgraphs of nodes $i = p_1$, $j = p_3$ correspondingly. (\ref{fig:fork1c}) represents all events. $j$ is about to create event $x$ (marked by purple) on receipt of a message from $i$. Node $i$ compares its last index array $(4, 4, 6, -)$ with $j$'s array $(-, 1, 6, 3)$ and sends to $j$ events $e_0, e_1$ among others. Node $j$ sees that digests $(h_0, id_{p}, 4)$ in $e_0$ and $(h_1, id_{p}, 6)$ in $e_1$ contain different hashes from those of events $c_0$ and $c_2$ correspondingly. This gives $j$ the idea that either $p$ or $i$ is cheating. Thus $j$ sends its compressed fork information $(h(a_0), h(c_2))$ to $i$ expecting $i$ to send its part of $p$'s fork. In case $i$ is honest it will send chain of events $a_1, a_2, b_0, b_1, b_2$ to $j$ and $j$ will get all $p$'s events known by $i$ and proof that $p$ forked. In case $i$ is faulty node $j$ will have a proof of protocol violation by $i$ since all $i$'s messages are signed by $i$.       
 }
 \label{fig:fork1}
\end{figure}

An example illustrating second part of this case is given on Figure \ref{fig:fork1}. 

\subsection{Cases 2 and 3: Node $j$ knows about some forks, $i$ either knows or doesn't know about any forks} 
In this case $j$ starts by sending to $i$ $\mathbf{lia}_j$ and all $\mathbf{cfi}_p$ where $p$ runs through all forking nodes known to $j$. Note that $j$ puts in $\mathbf{lia}_j$ indices of latest supported events created by forking nodes. 

In its turn, node $i$ acts almost as in the first case comparing $\mathbf{lia}_i$ with $\mathbf{lia}_j$ and sending to $j$ not only events with big enough indices but also \textit{tree branches} corresponding to $i$'s compressed fork information tuples that are absent among the ones it got from $j$. 

After getting this events node $j$ as in first case collects all invalid pairs of events $e, e'$ where $e$ contains a digest $(h, id_p, ind)$ and $e'$ has a digest $(h'\neq h, id_p, ind)$ for some node $p$. If there are no such pairs communication is successfully finished. 

\begin{figure}
 \begin{minipage}{.32\linewidth}
  \centering
  \subfloat[]{\label{fig:fork2a}\includegraphics[scale=.7]{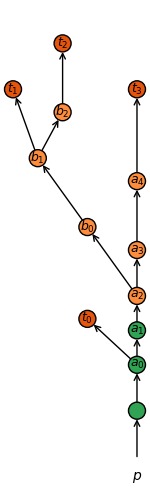}}
 \end{minipage}
 \begin{minipage}{.32\linewidth}
  \centering
  \subfloat[]{\label{fig:fork2b}\includegraphics[scale=.7]{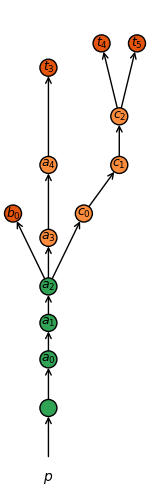}}
 \end{minipage}
 \begin{minipage}{.32\linewidth}
  \centering
  \subfloat[]{\label{fig:fork2c}\includegraphics[scale=.7]{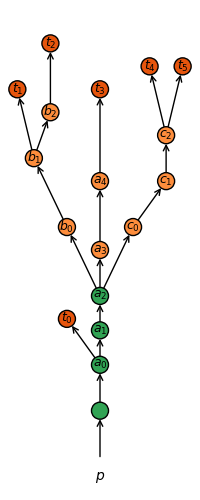}}
 \end{minipage}\par\medskip
 \caption{Example of faulty node $p$ event tree structure exchange in cases 2, 3. Figures (\ref{fig:fork2a}), (\ref{fig:fork2b}) represent nodes $i$ and $j$ knowledge about $p$'s event tree correspondingly.  (\ref{fig:fork2c}) is the whole $p$'s event tree. By green color we mark supported events in the corresponding hashgraphs. By our protocol $j$ sends index of $a_2$ in $\mathbf{lia}_j$ and compressed fork information $\mathbf{cfi}_p$, which is $(h(a_2), h(b_0)), (h(a_2), h(t_3)), (h(a_2), h(t_4)), (h(a_2), h(t_5))$ to $i$. As it was mentioned in remark \ref{more_cfi} this compressed fork information can be reduced to $(h(a_2), [h(b_0), h(t_3), h(t_4), h(t_5)])$. On receipt $i$ compares $\mathbf{lia}_j$ and $j$'s $\mathbf{cfi}_p$ with $i$'s knowledge about $p$'s event tree and sends back events $t_0, a_2, b_0, b_1, t_1, b_2, t_2$, which constitute missing by $j$ tree branches. In such a way $j$ knows all $p$'s events node $i$ is aware of.}
 \label{fig:fork2}
\end{figure}

Otherwise, to get all missing events it suffices for $j$ to send $i$ all compressed fork information about such suspicious nodes $p$. Node $i$ in its turn sends to $j$ all missing tree branches and communication is considered as successfully finished. We illustrate a fork information exchange by example on Figure \ref{fig:fork2}. 

\subsection{Case 4: Node $j$ is not aware of any forks but $i$ does}
Here $j$ starts by sending $\mathbf{lia}_j$ to $i$  and afterwards $i$ and $j$ act as in the previous cases 2 and 3.

\subsection{Remarks}
One can see that when there are any known forks in the system, nodes start to exchange their compressed fork information. If no new forks appear then there is no need to resend already sent $\mathbf{cfi}$. To achieve this each node can store fork information it sent to each other node and evade resending the same information to the same node again. 

One can also include in $\mathbf{cfi}$ not only branch tip hashes but also their graph distances from the root events in forking node event tree. These distances coincide with event index in case when tree is a sequence. Such additional information allows to save bandwidth in cases when one node knows a little bit more about certain tree branches than the other one.

\end{appendices}

\bibliographystyle{plain}
\bibliography{mybibdatabase.bib}

\end{document}